\documentclass[amsmath,amssymb,showpacs]{revtex4}
\usepackage{amsmath,amsfonts,amssymb,latexsym,graphicx,youngtab,extarrows}
\newtheorem{thm}{Theorem}
\newtheorem{lemma}{Lemma}

\newtheorem{rema}{Remark}
\newtheorem{corollary}{Corollary}

\renewcommand{\theequation}{\arabic{equation}}
\newcommand{\ch}{ {\large{\bf\ \  CHANGE !!!\ \ }} }

\newcommand{\dis}{\displaystyle}

\newcommand{\bequ}{\begin{equation}}
\newcommand{\eequ}{\end{equation}}
\newcommand{\barr}{\begin{array}}
\newcommand{\earr}{\end{array}}
\newcommand{\bea}{\begin {eqnarray}}
\newcommand{\eea}{\end {eqnarray}}
\newcommand{\lb}{\label}

\newcommand{\qed}{\hfill \rule{2.25mm}{2.25mm}\vspace{.15cm}}

\renewcommand{\Re}{{\cal R}{\mathrm e}\:}
\newcommand{\tr}{\mathrm Tr}
\begin{document}
% ====================================
\let\la=\lambda
\def \Z {\mathbb{Z}}
\def \Zt {\mathbb{Z}_o^4}
\def \R {\mathbb{R}}
\def \C {\mathbb{C}}
\def \La {\Lambda}
\def \ka {\kappa}
\def \vphi {\varphi}
\def \Zd {\Z ^d}
% ====================================
% **************************************************
% **************************************************
%\begin{frontmatter}
\title{On Thermodynamic and Ultraviolet Stability of Bosonic Lattice QCD Models in Euclidean Spacetime Dimensions $d=2,3,4$}
\author{Paulo A. Faria da Veiga}\email{veiga@icmc.usp.br}
\author{Michael O'Carroll}\email{michaelocarroll@gmail.com}
\affiliation{Departamento de Matem\'atica Aplicada e Estat\'{\i}stica - ICMC, USP-S\~ao Carlos,\\C.P. 668, 13560-970 S\~ao Carlos SP, Brazil}
% **************************************************
% **************************************************
% **************************************************
% **************************************************
% **************************************************
% **************************************************
% *******DATA*******DATA******DATA*****DATA*******DATA*****  DATA
\date{May 02, 2020%{\bf\Large DRAFT VERSION}
\vspace{.5cm}}
% **************************************************
% **************************************************
% **************************************************
% **************************************************
%\end{frontmatter}
% **************************************************
% **************************************************
% **************************************************
%
%
%
%
% **************************************************
\begin{abstract}
Using an imaginary-time, functional integral formulation, we prove thermodynamic and ultraviolet stable stability bounds for the relativistic local gauge-invariant lattice scalar QCD quantum model, with multiflavored real or complex scalar Bose matter fields replacing the usual fermionic (anti)quarks. We consider a local gauge-invariant gauge model with a compact and connected gauge Lie group $\mathcal G$. For concreteness, we concentrate on $\mathcal G={\mathrm U}(N)$, ${\mathrm SU}(N)$. Let $d(N)=N^2,(N^2-1)$ denote their Lie algebra dimensions, respectively. We start with the model defined on a finite hypercubic lattice $\Lambda\subset a\mathbb Z^d$, $d=2,3,4$, $a\in(0,1]$, with $L\in\mathbb N$ sites on a side, $\Lambda_s=L^d$ sites, and with free boundary conditions. The model action is a sum of a minimally coupled Bose-gauge part, which is quadratic in the Bose fields, and a Wilson pure-gauge plaquette term. We employ a priori local, scaled scalar multiflavor Bose fields, i.e. Bose fields with an $a$-dependent type of field-strength renormalization defining a non-canonical scaling. The Wilson pure-gauge action is a sum over gauge-invariant pointwise positive plaquette actions with a pre-factor $(a^{d-4}/g^2)$, and we take the gauge coupling $g^2$ in $(0,g_0^2]$, $0<g_0<\infty$. To eliminate the excess number of gauge bond variables, due to local gauge invariance, sometimes we use an enhanced temporal gauge so that there are $\Lambda_r\simeq (d-1)\Lambda_s$, for $L\gg 1$, retained bond variables. The value of the partition function is not altered since there are no loops involved and only the bonds of a maximal tree in $\Lambda$ are set equal to the identity. Considering the finite-lattice original, unscaled partition function $Z^u_{\Lambda,a}\equiv  Z^u_{\Lambda,a,\kappa_u^2,m_u,g^2,d}$, where $\kappa_u^2>0$ is the unscaled (bare) hopping parameter and $m_u\geq 0$ are the boson fields bare masses, and letting $s_B\equiv [a^{d-2}(m_u^2a^2+2d\kappa_u^2)]^{1/2}$ and $s_Y\equiv a^{(d-4)/2}/g$, we show that the scaled partition function $Z_{\Lambda,a}=s_B^{N\Lambda_s}s_Y^{d(N)\Lambda_r} Z^u_{\Lambda,a}\,$  satisfies the thermodynamic and ultraviolet stable stability bounds
$$e^{c_\ell d(N)\Lambda_s}\leq Z_{\Lambda,a}\leq e^{c_ud(N)\Lambda_s}\,,$$
with finite constants $c_\ell,\,c_u\in\mathbb R$ independent of the lattice size $L$ of $\Lambda$ and the lattice spacing $a$. The gauge fixing is used only to prove the lower bound on $Z^u_{\Lambda,a}$. In these stability bounds, we have extracted and isolated in $Z^u_{\Lambda,a}$ the dependence on $\Lambda$ and the  {\em exact} singular behavior of the finite lattice free energy in the continuum limit $a\searrow 0$. For the normalized finite-lattice free energy $f_\Lambda^n=[d(N)\,\Lambda_s]^{-1}\,\ln Z_{\Lambda,a}$, the Bolzano-Weierstrass theorem ensures that a finite thermodynamic limit ($\Lambda\nearrow a\mathbb Z^d$) exists for $f_\Lambda^n$, at least in the sense of subsequences. Subsequently, a continuum limit $a\searrow 0$ also exists, at least in the subsequential sense, and defines a model normalized free energy $f^n$. The existence of $f^n$ is the only question we examine here! Our stability bounds on $Z_{\Lambda,a}$ are proved by showing partial stability bounds separately for the scaled Wilson pure-gauge partition function $Z_{Y,\Lambda,a}\equiv  Z_{Y,\Lambda,a,g^2,d}$ and the scaled Bose-gauge partition function $Z_{B,\Lambda,a}(\tilde g)\equiv  Z_{B,\Lambda,a,\kappa_u^2,g^2,m_u,d}(\tilde g)$ with fixed gauge coupling $g^2$ and postponed integration over the gauge fields $\tilde g$. A simple bound relates the bounds on these two partial partition functions with bounds on the complete scaled partition function $Z_{\Lambda,a}$. We prove that the scaled pure-gauge partition function $Z_{Y,\Lambda,a}= s_Y^{d(N) \Lambda_r}Z^w_{Y,\Lambda,a}$ verifies thermodynamic and ultraviolet stable stability bounds. A new quadratic upper bound on the Wilson plaquette action in the gluon fields is obtained. Similarly, using a new method which bypasses diamagnetic inequality considerations, we prove that the scaled Bose-matter partition function $Z_{B,\Lambda,a}(\tilde g)=s_B^{N\Lambda_s}Z^u_{B,\Lambda,a}(\tilde g)$ is also thermodynamic and ultraviolet stable. The bounds are uniform in the bond variables $\tilde g$. Our methods extend to treat more general lattices and other compact, connected Lie gauge groups.
\end{abstract}
\maketitle
%%%%%%%%%%%%%%%%%%%%%%%%%%%%%%&&&&&&&&&&&&&&&&&&&&&&&%%%%%%%%%%%%%%%%%%%%%%%%%%%%%
\section{Introduction and Results} \lb{sec1}
%%%%%%%%%%%%%%%%%%%%%%%%%%%%%%&&&&&&&&&&&&&&&&&&&&&&&%%%%%%%%%%%%%%%%%%%%%%%%%%%%%
To prove the existence of QFT models in spacetime dimension $d=4$ is still one of the most fundamental and challenging open problems in physics. Using Euclidean spacetime to analyze this question, and show the existence of a model in the thermodynamic and the continuum limits, one approach is to derive stability bounds for its partition function and prove the existence of the free energy \cite{Rue}. Then, once we show bounds to the generating functions of correlations, we can obtain correlations and, from them, we can derive properties for the limiting models, such as to obtain the corresponding energy-momentum spectrum.

During this process, if Osterwalder-Schrader positivity \cite{GJ,JJ} is verified and an underlying physical, quantum mechanical Hilbert space $\mathcal H$, is constructed, the mutually commuting, self-adjoint positive-energy and spatial-momentum operators can be defined. Applying the spectral theorem to these operators allows us to establish spectral representations for correlations which provide, via a Feynman-Kac type formula, a rigorous connection between the singularities in complex momentum space of the Fourier transforms of some correlations with points in the E-M spectrum. Last, once all the Osterwalder-Schrader axioms are verified, we can also verify Wightman axioms for the corresponding model in the Minkowski spacetime, and eventually obtain a physically interesting relativistic QFT in Minkowski spacetime \cite{Wei,Banks,Zee,Gat,GJ,Riv,Simon1,Dim1,Sei,Summers}.

The best candidate to fulfill the above requirements in $d=4$ is the ${\mathrm SU}(3)_c$ local gauge-invariant model of Quantum Chromodynamics (QCD), defined with three-colored fermionic quarks and antiquarks.

As an intermediate step to deal with this problem, here we analyze local abelian and non-abelian gauge models coupled with bosonic scalar matter fields instead of (anti)quarks. It is known that replacing the fermionic matter fields by Bose fields introduces mathematical problems not present in Fermi models. This is due to the fact that boson field operators are unbounded, as opposed to the case of Fermi fields which, due to the Pauli principle, are bounded operators. However, as shown below, we can handle this difficulty and prove thermodynamic and ultraviolet stable stability bound and the existence of the free energy for bosonic QCD in $d=2,3,4$. Also, from a physical viewpoint, our bosonic QCD model is relevant since it can be used to treat e.g. effective interactions between hadrons. Besides, the methods established in \cite{MP,M,MP2} work well for both Bose and Fermi lattice gauge models.

The pure-gauge term was neglected in Refs.\cite{MP,M,MP2}, and was treated independently in \cite{YM}, where we obtained thermodynamic and ultraviolet stable stability bounds for the pure-gauge, Yang-Mills model defined on the lattice and using the Wilson plaquette action. The same (noncanonical) scaling ideas are behind our method in \cite{YM}, which is simple, direct, and does not make use of sophisticated multiscale renormalization group analysis employed in the important work carried out by Balaban (see \cite{Bal,Bal2} and references therein). In \cite{BFS}, abelian gauge Bose matter models were considered in $d=2$. For $d=3$, the $\mathrm U(1)$ abelian case was considered in \cite{King} and, more recently, in Ref. \cite{Dim}.

Considering Fermi-gauge models, like QCD, although in \cite{MP,MP2} we showed an upper stability bound for the partition function and also bounded generating functions, we have not yet attacked the question of proving a lower stability bound. This is an important open question.

In this work, we use the methods of Refs. \cite{M,YM} to prove upper and lower thermodynamic and ultraviolet stable stability bounds for a scaled partition function for lattice bosonic QCD models in Euclidean dimensions $d=2,3,4$. Both the upper and lower stability bounds imply the same singularity of the free energy as the lattice spacing $a\searrow 0$, i.e. in the continuum limit. Then, by the Bolzano-Weierstrass theorem, we prove the existence of the thermodynamic and continuum  corresponding normalized free energy for the models, at least in the subsequential sense.

Our starting point is the finite-lattice Bose-gauge model with unscaled partition function
\bequ\lb{Z}
Z^u_{\Lambda,a}\equiv  Z^u_{\Lambda,a,\kappa_u^2,m_u,g^2,d}=\int\,e^{-[S^u_{\Lambda,a}(\tilde\phi^u,\tilde g)]}\;d\tilde\phi^u\,d\tilde g\,,
\eequ
where the action is \bequ\lb{action}
S^u_{\Lambda,a}(\tilde\phi^u,\tilde g)=S^u_{B,\Lambda,a}(\tilde\phi^u,\tilde g)+S^w_{Y,\Lambda,a}(\tilde g)\,.\eequ
Here, $\tilde\phi^u$ denotes a collection of real or complex scalar fields and $\tilde g$ are the gauge variables taking values in a compact and connected Lie gauge group $\mathcal G$. For concreteness, we concentrate our attention on $\mathcal G\,=\,{\mathrm U}(N),\;{\mathrm SU}(N)$ and denote by $d(N)=N^2,N^2-1$ their Lie algebra dimensions, respectively.

The Bose-gauge action in Eq.(\ref{action}) is given by the lattice minimally coupled approximation with lattice sites denoted by $x$. Formally, it is given by
\bequ\lb{suM}\barr{lll}
A^u_{B,a}({\tilde{\phi}^u})&=&\frac{\kappa_u^2}2\,a^{d-2}\,\dis\sum_{x,\alpha,\gamma,\xi,f}\:\sum_{\mu=0}^{d-1}\,\left[\left(\phi^u_{\alpha,f}(x^+_\mu)-\left(g_{b_\mu(x)}\right)_{\alpha\gamma}\phi^u_{\gamma,f}(x)\right)^{cc}\,\left(\phi^u_{\alpha,f}(x^+_\mu)-\left(g_{b_\mu(x)}\right)_{\alpha\xi}\phi^u_{\xi,f}(x)\right)\right]\vspace{2mm}\\&&+\,\frac12\,m_u^2a^d\dis\sum_{x,\alpha,f}\,|\phi^u_{\alpha,f}(x)|^2\,,\earr
\eequ
where the first sum means the sum over lattice sites $x=(x^0,x^1,\ldots,x^{d-1})$, with $x^0$ denoting the temporal coordinate, $\alpha,\gamma,\xi=1,\ldots,N$ are gauge group indices (color), $f=1,\ldots,N_f$, $N_f\in\mathbb N$ is the flavor number and $cc$ means complex conjugation. Here, $x^+_\mu\,\equiv\,x\,+\,ae^\mu$, where $e^\mu$ is the unitary vector of the $\mu=0,1,\ldots,(d-1)$ direction and $b_\mu(x)$ denotes a directed lattice bond $\langle x,x^+_\mu\rangle$, originating at site $x$ and ending at site $x^+_\mu$. For simplicity, below we take $N_f=1$ and suppress the color, gauge index $\alpha$. The general case with $N_f$ flavors may be easily recovered from our analysis.

Here, we work with a finite hypercubic lattice $\Lambda\subset(a\mathbb Z)^d$, with $L\in\mathbb N$ sites on a side, and we denote the number of sites in $\Lambda$ by $\Lambda_s=L^d$. Many of our results extend to more general lattices. Based on Eq. (\ref{suM}), the action $S^u_{B,\Lambda,a}(\tilde\phi^u,\tilde g)$ of Eq. (\ref{action}), with free boundary conditions, is defined by
\bequ\lb{actionB}
e^{-S^u_{B,\Lambda,a}(\tilde \phi,\tilde g)}\,=\,\prod_{b_\mu(x)} F^u_{b_\mu(x)}\,\prod_{x\in\Lambda}\, \exp\left[-\frac12 \left(2d\kappa_u^2a^{d-2}+a^dm_u^2\right)\,|\phi^u(x)|^2\right]\,,
\eequ
where, for free boundary conditions and with the condition that both sites $x,\,x^+_\mu\in\Lambda$, we have the {\em bond factor}
\bequ\lb{weight}
F^u_{b_\mu(x)}=\exp\,\left\{\frac{\kappa_u^2a^{d-2}}2\,\left[\bar\phi^u(x)
g_{b_\mu(x)} \phi^u(x^+_\mu)
+\bar\phi^u(x^+_\mu)
g^{-1}_{b_\mu(x)} \phi^u(x)\right]\right\}\,.
\eequ
In the sequel, we will need to use the corresponding unscaled Bose-gauge partition function with an arbitrary gauge field configuration and postponed gauge integral, namely,
\bequ\lb{ZuM}
Z^u_{B,\Lambda,a}(\tilde g)\,=\,\int\,e^{-S^u_{B,\Lambda,a}(\tilde\phi,\tilde g)}\;d\tilde\phi\,.
\eequ

Expanding the bond factor in powers of $\kappa^2$, and truncating the series at order $\kappa^{2\eta}$, defines what we call the $\eta$-truncated model. This truncated model also satisfies Osterwalder-Schrader positivity and mimics the Fermi-gauge part of fermionic QCD, where the truncation occurs naturally for fermions due to the Pauli exclusion principle.

For further considerations, we observe that we can also define the model with periodic boundary conditions, which we do not use here, by adding links to the bond factor connecting hypercubic lattice final to initial points in each coordinate direction. For the free and periodic boundary conditions, our Bose-gauge models satisfy Osterwalder-Schrader positivity for $\kappa_u^2,\,m_u^2>0$.
Besides, we emphasize that the exponent in Eq. (\ref{weight}) is real and, of course, the free Bose field case is obtained from the above action by setting all $g$ equal to the identity matrix.

The (spin zero) Bose fields $\tilde{\phi}(x)\equiv \phi_{c,f}(x)$ are defined at the lattice sites $x$, and are taken in the fundamental representation of $\mathcal G$. For each directed lattice bond $b_\mu(x)\,=\,\langle x,x^+_\mu\rangle$ in the $\mu$-direction, we assign a gauge bond variable $g_{b_\mu(x)}$. This gauge bond variable is an $N\times N$ unitary matrix taken as an element of an irreducible representation of the gauge group $\mathcal G$.

The pure-gauge, Yang-Mills action $S^w_{Y,\Lambda,a}(\tilde g)$ in Eq. (\ref{action}) is the Wilson plaquette action (see Refs. \cite{Wil} and \cite{Sei})
\bequ\lb{YMaction}
S^w_{Y,\Lambda,a}\,=\,\dfrac {a^{d-4}}{g^2}\,\sum_p\,\mathcal A_p\,\equiv\,\dfrac {a^{d-4}}{g^2}\,\dis\sum_p\,\|1-g_{p}\|^2_{H-S}\,,
\eequ
with $g^2\in(0,g_0^2]$, $0<g_0^2<\infty$.
Let $p_{\mu\nu}(x)$ denote the plaquette (oriented minimal square with four lattice bonds) in the $\mu\,-\,\nu$ plane, with $\mu<\nu$; $\mu,\nu=0,1,\ldots,(d-1)$, and vertices at $x$, $x_\mu^+$, $x_\mu^++ae^\nu$, $x_\nu^+$, then $$g_{p}\,=\,g_1g_2g_3g_4\,,$$ where $g_1=g_{\langle x,x^{+}_\mu\rangle}$, $g_2=g_{\langle x^{+}_\mu,x^{+}_\mu+ae^\nu\rangle}$, $g_3=g^{-1}_{\langle x^{+}_\nu,x^{+}_\nu+ae^\mu\rangle}$ and $g_4=g^{-1}_{\langle x,x^{+}_\nu\rangle}$. Here, $g^{-1}$ denotes the inverse gauge group element and, for a square matrix $M$, we have the Hilbert-Schmidt norm squared
$$
\|M\|^2_{H-S}\,\equiv\,Tr(M^\dagger M)\,\geq\,0\,,
$$
where $Tr$ is the trace and $^\dagger$ denotes the adjoint. In Eq. (\ref{YMaction}), $\sum_p$ runs over all plaquettes in the hypercubic lattice $\Lambda\subset a\mathbb Z^d$. We can also define, for a hypercubic lattice, the model with periodic boundary conditions by considering additional plaquettes joining the left most plaquettes to the right most plaquettes in each coordinate direction plus {\em corner} plaquettes.

Note that each plaquette action in $S^w_{Y,\Lambda,a}$ is pointwise positive  and this property is global, such as it holds for the whole group $\mathcal G$. Also, as the gauge group $\mathcal G$ is unitary then $g_{p}$ is also a unitary matrix and can be written as $g_{p}\,=\,\exp[iX]$, with $X$ self-adjoint. As $\mathcal G$ is compact and connected, the exponential map is onto \cite{Simon2,Hall,Far}. Also, $$\|1-g_p\|^2_{H-S}=2\Re {\mathrm Tr}(1-g_p)\,=\,{\mathrm Tr}[(1-e^{iX})^\dagger(1-e^{iX})]\,=\,2\,{\mathrm Tr}(1\,-\,\cos X)\,.$$

If $b$ is identified with the bond $b_\mu(x)$ then, formally, $g_b\,=\,e^{iagA_b}$, where $A_b$ is identified with the $d(N)$ gauge potentials or physical gluon fields $A^c_\mu(x)$. The fields $A_b\,=\,\sum_{c=1,\ldots,d(N)} A_b^c \theta_c$, where the $\theta_c$, $c=1,\ldots,d(N)$, are $N\times N$ self-adjoint matrices which form a basis for the Lie algebra for the group $\mathcal G$. The index $c$ fixes the gluon color, and we normalize the Lie algebra generators such that ${\mathrm Tr}\theta_\alpha\theta_\beta=\delta_{\alpha\beta}$, with a Kronecker delta, so that $A_b^\alpha\,=\,\tr A_b\theta_\alpha$.

The unscaled partition function associated with the free-boundary action $S^w_{Y,\Lambda,a}$ is defined by
\bequ\lb{Zuw}
Z^w_{Y,\Lambda,a}\,=\,\dis\int\,e^{-S^w_{Y\Lambda,a}}\,d\tilde g\,.
\eequ
Here, the gauge field measure $d\tilde g$ is the product over bonds $b_{\mu}(x)$ of normalized to one Haar measures $d\sigma(g_{b_\mu(x)})$ on the gauge group $\mathcal G$ \cite{Simon2,Hall,Far}.

Using the Baker-Campbell-Hausdorff formula, formally, in Ref. \cite{Gat} it is shown that the Wilson plaquette action of Eq. (\ref{YMaction}) approximates the continuum smooth field classical Yang-Mills action as $a\searrow 0$ (see below). Similarly, the Bose-gauge action of Eq. (\ref{suM}) approximates the classical Bose-gauge field minimal coupling as $a\searrow 0$. Namely, with $g_{b_\mu(x)}\,=\,\exp[iagA_\mu(x)]$, $A_\mu(x)\,=\,\sum_{\alpha=1,\ldots,d(N)}A_\mu^\alpha(x)\theta_\alpha$ and the covariant derivative $D_\mu\,\equiv\,\partial_\mu\,-\,igA_\mu(x)$, the action in the continuum spacetime is given by
$$
\dis\int\,\left\{\left(\kappa_u^2/2 \right)\,[D_\mu\phi(x)]^{cc}\,D_\mu\phi(x)\, +\,\dfrac12\, m_u^2\,|\phi(x)|^2\right\}\,d^dx\,.
$$

In Eq. (\ref{Z}), writing the complex scalar field as $\phi_{c,f}(x)=\phi_{R,c,f}(x)+i\phi_{I,c,f}(x)$, then the measure $d\tilde{\phi}$ is a product of Lebesgue measures
$$
d\tilde{\phi}\,=\,\prod_{x\in\Lambda}\,\prod_{c=1,\ldots,N}\,\prod_{f=1,\ldots,N_f} \,\left[\,\left(\frac1{\sqrt{2\pi}}\,d\phi_{R,c,f}(x)\right)\:\left(\frac1{\sqrt{2\pi}}\,d\phi_{I,c,f}(x)\right)\,\right]\,.
$$
For the real Bose field case, we set $\phi_{I,c,f}=0$ and neglect the measure in the second parenthesis on the right-hand-side above.

The actions  $S^u_{B,\Lambda,a}(\tilde\phi^u,\tilde g)$, $S^w_{Y,\Lambda,a}(\tilde g)$, and then $S^u_{\Lambda,a}(\tilde\phi^u,\tilde g)$, as well as the partition function $Z^u_{\Lambda,a}$ of Eq. (\ref{Z}) are invariant under the usual local gauge transformations. Namely, the group $\otimes_{x\in\Lambda}\, \mathcal G$, with elements $\prod_x r_x$,  maps $\phi_{c,f}(x)$ to $r_x\phi_{c,f}(x)$, $\phi^\dagger_{c,f}(x)$ to $r^{-1}_x\phi^\dagger_{c,f}(x)$ and acts on bond variables  mapping $g_{b_{\mu(x)}}$ to $r_x g_{b_{\mu(x)}} r^{-1}_{x+ae^\mu}$.

For simplicity, here, we will treat explicitly only the case $\mathcal G={\mathrm U}(N)$, but we also show how to extend our analysis to the case $\mathcal G={\mathrm SU}(N)$. The extension to the multiflavored case $N_f>1$ is obvious.

Due to local gauge invariance, there is an excess of bond variables. The excess variables can be eliminated (or, say, {\em gauged away}) by using the {\sl enhanced temporal gauge}. In this gauge, the temporal bond variables and certain specified bond variables on the boundary of $\Lambda$ are set to the identity matrix in the action. The corresponding gauge integrals are dropped. In $d=2,3,4$, the gauged away bonds do not form loops. We denote the number of remaining or {\em retained} gauge variables by $\Lambda_r\equiv\Lambda_r(d)$ which has the values $(L-1)^2$, $[(2L+1)(L-1)^2]$, $[(3L^3-L^2-L-1)(L-1)]$, for $d=2,3,4$, respectively. Clearly $\Lambda_r\simeq (d-1)L^d$, for $L\gg 1$, and $\Lambda_r\nearrow\infty$ as $\Lambda\nearrow a\mathbb Z^d$. Letting $\Lambda_p$ be the number of distinct plaquettes in the lattice $\Lambda$, we emphasize that $\Lambda_p=\Lambda_r$ for $d=2$, but not for $d=3,4$.

To be more precise, if we identify the sites of the $\mu$-th coordinate with $1,2,\ldots,L$, the enhanced temporal gauge is defined by setting, in the lattice $\Lambda$, the following bond variables to $1$. First, for any $d=2,3,4$, we take $g_{b_0(x)}=1$. For $d=2$, take also $g_{b_1(x^0=1,x^1)}=1$. For $d=3$, set also $g_{b_1(x^0=1,x^1,x^2)}=1$ and $g_{b_2(x^0=1,x^1=1,x^2)}=1$. Similarly, for $d=4$, take also $g_{b_1(x^0=1,x^1,x^2,x^3)}=1$, $g_{b_2(x^0=1,x^1=1,x^2,x^3)}=1$ and $g_{b_3(x^0=1,x^1=1,x^2=1,x^3)}=1$. For $d=2$, the gauged away bond variables form a comb with the teeth along the temporal direction, and the open end at the maximum value of $x^0$. For $d=3$, the gauged away bonds can be visualized as forming a scrub brush with bristles along the $x^0$ direction and the grip forming a comb. In $d=2,3,4$, the gauged away bonds do not form loops, and note that, fixing the enhanced temporal gauge, the unretained or gauge away variables are associated with bonds in the hypercubic lattice $\Lambda$ which form a maximal tree, so that by adding any other bond to it we form a closed loop.

We assume the enhanced temporal gauge fixing only to prove the lower stability bound of Eq. (\ref{stab1}). In this case, for simplicity, we make an abuse of notation and keep using the same notation as before everywhere.

To prove the upper and lower stability bounds, our goal is to multiplicatively transform the unscaled partition function $Z^u_{\Lambda,a}$ so that the scaled partition function $Z_{\Lambda,a}$ satisfies a thermodynamic and ultraviolet stable stability bound
\bequ\lb{stab1}e^{c_\ell d(N)\Lambda_s}\leq Z_{\Lambda,a}\leq e^{c_ud(N)\Lambda_s}\,,
\eequ
with finite $c_\ell,\,c_u\in\mathbb R$ independent of $L$ (or $\Lambda$), the lattice spacing $a\in(0,1]$ and $g^2\in(0,g_0^2]$, $g_0<\infty$.

By the Bolzano-Weierstrass theorem, an important consequence of the stability bounds of Eq. (\ref{stab1}) is that the normalized finite lattice free energy per site
\bequ\lb{fn}
f^n_{\Lambda,a}\,=\,\dfrac 1{\Lambda_s}\, \ln Z_{\Lambda,a}\,,
\eequ
satisfies the bounds $-\infty<c_\ell\leq f^n_{\Lambda,a}\leq c_u<\infty$, such that we can establish the existence of a finite thermodynamic limit $\Lambda\nearrow a\mathbb Z^d$ and, subsequently, a finite continuum limit $a\searrow 0$ of the free energy, at least in the sense of subsequences.

We now show that to determine a scaled partition function $Z_{\Lambda,a}$ verifying the stability bounds given in Eq. (\ref{stab1}), it suffices to consider {\em separately} a scaled Bose-gauge partition function $Z_{B,\Lambda,a}(\tilde g)$, with integrated Bose fields and fixed arbitrary gauge fields $\tilde g$, and a scaled pure-gauge partition function $Z_{Y,\Lambda,a}$. To explain why we can proceed in this way is justified by the simple, generic bound applied at the beginning of our analysis (recall $\Lambda$ is a finite set and $\mathcal G$ is compact!)
\bequ\lb{principal}
\left[{\mathrm min}_{\tilde g}\left(Z^u_{B,\Lambda,a}\right)(\tilde g)\right]\,Z^w_{Y,\Lambda,a}\,\leq
Z^u_{\Lambda,a}\,=\,\int\,e^{-[S^u_B(\tilde\phi,\tilde g)+S^w_Y(\tilde g)]}\;d\tilde\phi\,d\tilde g\,\leq\,\left[{\mathrm max}_{\tilde g} \left(Z^u_{B,\Lambda,a}\right)(\tilde g)\right]\,Z^w_{Y,\Lambda,a}\,.
\eequ
where $Z^u_B(\tilde g)$ and $Z^w_{Y,\Lambda,a}$ are, respectively, defined in Eqs. (\ref{ZuM}) and (\ref{Zuw}).
\begin{rema}
Note that, in the bound of Eq. (\ref{principal}), we have the freedom to fix or not fix the gauge before doing it. In the former case, both $Z^u_B(\tilde g)$ and $Z^w_{Y,\Lambda,a}$ will inherit the previously fixed gauge. In the latter case, using the gauge invariance of $Z^w_{Y,\Lambda,a}$ itself, we may ulteriorly fix any gauge after applying the bound.
\end{rema}

For the Bose-gauge partition function with postponed integration over the gauge field, the thermodynamic and ultraviolet stable stability bound is achieved by passing from the unscaled Bose field $\phi^u(x)$ to the a priori scaled Bose field at each lattice point $x$ by the scaling transformation,
\bequ\lb{scB}
\phi(x)\,=\,s_B\,\phi^u(x)\qquad;\qquad s_B\equiv s_B(a)=[a^{d-2}(m_u^2a^2+2d\kappa_u^2)]^{1/2}\,.
\eequ
Of course this transformation is {\em not} to be confused with the canonical scaling! Formally, the infinite lattice scaled action is
\bequ\lb{sM}
A_{B,a}({\tilde{\phi}})\,=\,-\,\frac{\kappa^2}2\,\sum_{x}\sum_{\mu=0}^{d-1}\,\left[\bar\phi(x)g_{b_\mu(x)}\phi(x^+_\mu)+\bar\phi(x^+_\mu)g^{-1}_{b_\mu(x)}\phi(x)\right]\,+\,\frac12\,\sum_{x}\,|\phi(x)|^2\,,
\eequ
where the scaled hopping parameter to $\kappa^2$ is given by (with $u\,\equiv\,\kappa_u^2/m_u^2$)
\bequ\lb{hopbose}
\kappa^2\,=\,\left[2d\,+\,\left(\dfrac{m_ua}{\kappa_u}\right)^2\right]^{-1}\,=\, \dfrac{u}{a^2+2du}\,.%\leq\,\frac1{2d}\,.
\eequ
Clearly $\kappa^2>0$ satisfies the inequality, for $a\in(0,1]$, \bequ\lb{limkappa}[2d\,+\,(m_u/\kappa_u)^2]^{-1}\leq \kappa^2\leq (2d)^{-1}\,,\eequ and the unscaled Bose-gauge partition function $Z^u_{B,\Lambda,a}(\tilde g)$ is transformed into the free-boundary, unit Bose mass scaled partition function $Z_{B,\Lambda,a}(\tilde g)$ defined by Eq. (\ref{ZM}), with
$$
e^{-S_{B,\Lambda,a}(\tilde \phi,\tilde g)}\,=\,\prod_{b_\mu(x)} F_{b_\mu(x)}\,\prod_{x\in\Lambda} e^{-\frac12|\phi(x)|^2}\,,
$$
and, for free boundary conditions,
\bequ\lb{scaledweight}
F_{b_\mu(x)}=\exp\,\left\{\frac{\kappa^2}2\,\left[\bar\phi(x)
g_{b_\mu(x)} \phi(x^+_\mu)
+\bar\phi(x^+_\mu)g^{-1}_{b_\mu(x)} \phi(x)\right]\right\}\,.
\eequ
The corresponding scaled gauge-Bose partition function with postponed integration on the gauge fields is given by
\bequ\lb{ZM}
Z_{B,\Lambda,a}(\tilde g)\,=\,\int\,e^{-S_{B,\Lambda,a}(\tilde\phi,\tilde g)}\;d\tilde\phi\,.
\eequ
such that, for real fields,
\bequ\lb{ZuBZB}
Z_{B,\Lambda,a}(\tilde g)\,=\,s_B^{N\Lambda_s}\,Z^u_{B,\Lambda,a}(\tilde g)\,.
\eequ
For complex Bose fields, change $s_B$ to $s_B^2$. The scaling of Eq.(\ref{scB}) is seen to transform the unscaled field action of Eq. (\ref{actionB}) to the scaled field action of Eq. (\ref{sM}). The scaled action has coefficients which are non-vanishing and non-singular in $a\in(0,1]$, in contrast with the unscaled field action which has singular and vanishing coefficients as $a\searrow 0$. With this scaling, an $a$-dependent factor is extracted from the unscaled field partition function and a new scaled field partition function is defined, which obeys thermodynamic and ultraviolet stability bounds.

What happens to correlations under this scaling transformation? The real field unscaled normalized correlations, with fixed gauge fields $\tilde g$,
$$
C^{u,(n)}_{B,\Lambda,a}(\tilde g;x_1,\ldots,x_n)\,=\,\dfrac1{Z^u_{B,\Lambda,a}(\tilde g)}\,\int\,\prod_{1\leq\gamma\leq n}\phi^u(x_\gamma)\;\,e^{-S^u_{B,\Lambda,a}(\tilde\phi,\tilde g)}\;d\phi^u\,,
$$
% where $Z^u_{B,\Lambda,a}\,=\,\int\,Z^u_{B,\Lambda,a}(\tilde g)\,d\tilde g$ 
are related to the scaled field correlations
$$
C^{(n)}_{B,\Lambda,a}(\tilde g;x_1,\ldots,x_n)\,=\,\dfrac1{Z_{B,\Lambda,a}(\tilde g)}\,\int\,\prod_{1\leq\gamma\leq n}\phi(x_\gamma)\;\,e^{-S_{B,\Lambda,a}(\phi,\tilde g)}\;d\phi\,,
$$
%\ch with $Z_{B,\Lambda,a}\,=\,\int\,Z_{B,\Lambda,a}(\tilde g)\,d\tilde g$, 
by 
$$
C^{(n)}_{B,\Lambda,a}(\tilde g;x_1,\ldots,x_n)\,=\,[s_B(a)]^{n}\, C^{u,(n)}_{B,\Lambda,a}(\tilde g;x_1,\ldots,x_n)\,.
$$

For the case of the free field, the scaled correlations are not singular in $a\in(0,1]$, such as pointwise bounded, uniformly in $a\in(0,1]$.

Besides, the temporally infinite decay rate of the scaled and unscaled correlations are the same. This means the energy-momentum spectral properties of the original unscaled lattice model do not alter and are the same as those of the scaled lattice quantum fields model.

Also, as is well-known in many models, renormalization of the action parameters has to be performed to keep the energy-momentum spectrum from running to infinity as $a\searrow 0$. Using unscaled field perturbation theory, the terms can be infinite; but here, using scaled fields, they are not!

Furthermore, for large dimension $d$, $\kappa^2$ is small and polymer expansions can be applied successfully, rigorously and explicitly (see \cite{Sei,Simon3}), to obtain the scaling limit of correlations. 

What is the effect of the local scaling in the much studied models where the action is the sum of a free field and a local polynomial? For $d=2$, the small behavior of the scaling factor is trivial. For $d\geq 3$, the free field part of the action scales as defined above to produce finite coefficients $\kappa^2$ and 1/2 for the hopping parameter and mass terms, respectively. The local interaction polynomial of the e.g type $\lambda a^d\,\sum_x[\phi^u(x)]^4$ is transformed to $\lambda (a^d/s_B^4)\,\sum_x[\phi(x)]^4$. For small $a$, $(a^d/s_B^4)\simeq a^{4-d}$ which is nonsingular for $d=3,4$; singular for $d\geq 5$. Scaled fields can be used in the analysis of these models in $d=3,4$, and give rise to more regularity of correlations, as in Refs. \cite{M,GYM}. When fermions are present, the same property was confirmed in \cite{MP}. In the case of mass renormalization, typically, as $a\searrow 0$, with scaled fields, $\kappa^2$ remains small and finite; the bare mass parameter $m_u^2$ becomes negative but $(m_u^2a^2)$ remains finite. 

With this and having Eq. (\ref{scB}) in mind, by inserting the Bose field scaling factors $s_B$ in Eq. (\ref{principal}), we obtain
\bequ\lb{principal2}
\left[{\mathrm min}_{\tilde g}\left(Z_{B,\Lambda,a}\right)(\tilde g)\right]\,Z^w_{Y,\Lambda,a}\,\leq s_B^{N\Lambda_r}\,
Z^u_{\Lambda,a}\,\leq\,\left[{\mathrm max}_{\tilde g} \left(Z_{B,\Lambda,a}\right)(\tilde g)\right]\,Z^w_{Y,\Lambda,a}\,.
\eequ
Below, by introducing another multiplicative factor, we show how to define a scaled pure-gauge partition $Z_{Y,\Lambda,g}$. Doing this, in these bounds, we have bounds for the complete Bose-gauge model scaled partition function $Z_{\Lambda,a}$.

Our first result is a thermodynamic and ultraviolet stable stability bounds on the scaled Bose-gauge finite-volume partition function $Z_{B,\Lambda,a}(\tilde g)$ given in the theorem below. The theorem applies for the case of a hypercubic lattice and free boundary conditions.
\begin{thm}\lb{thm1}
For an arbitrary gauge configuration $\tilde g$, the scaled Bose-gauge partition function $Z_{B,\Lambda,a}(\tilde g)$ satisfies the stability bounds
\bequ\lb{stab11}e^{c_{B,\ell}\Lambda_s}\leq Z_{B,\Lambda,a}(\tilde g)\leq e^{c_{B,u}\Lambda_s}\,,\eequ
with finite $c_{B,\ell},\,c_{B,u}\in\mathbb R$ independent of the size $L$ of the lattice $\Lambda$, the lattice spacing $a\in(0,1]$ and $g^2\in(0,g_0^2]$, $g_0<\infty$. The massless case $m_u=0$ is allowed with $\kappa^2=(2d)^{-1}$.
\end{thm}
\begin{rema}
The fact the bounds of Theorem \ref{thm1} do not depend on the bond variables $g$ is important so that we can easily take the maximum and the minimum in the generic bounds of Eq. (\ref{principal}).
\end{rema}
\begin{rema}
The scaled field correlations have the same temporal decay rates as the unscaled correlations for a temporally infinite lattice. Surprisingly, the scaled correlations are pointwise bounded for $d=3,4$ (see Refs. \cite{M,MP,MP2}). Using scaled fields yields more regularity of correlations.
\end{rema}
\begin{rema}
As it can be checked in the proof of Theorem \ref{thm1}, given in section \ref{sec2}, the proof still holds for pure imaginary hopping parameters $\kappa$, so that $-(2d)^{-1}\leq \kappa^2\leq (2d)^{-1}$. Negative real $\kappa^2$ values correspond to the antiferromagnetic case for which stability is also obtained for the scaled free field.
\end{rema}
\begin{rema}
Using diamagnetic inequalities and hypercubic lattices, in Ref. \cite{BFS}, it is shown that $Z_{B,\Lambda,a}(\tilde g)$ is bounded from above by the free Bose-field partition function, which is obtained by setting all gauge variables to the identity. For a proof of diamagnetic inequalities when more general lattices are considered, see \cite{M}. In these approaches, an upper stability bound is obtained by showing an upper stability bound for the free partition function. As it can be checked in the sequel, our method is new, direct and bypasses diamagnetic inequality considerations used in \cite{BFS}. The upper bound given in Theorem \ref{thm1} is obtained by applying H\"older's inequality to bound the Bose-gauge partition function by a product of partition functions of one-dimensional chains. Furthermore, the partition function of each chain is bounded by a product of bounds (one factor for each bond of the chain) of a single bond transfer matrix.
\end{rema}
\begin{rema}
The use of free boundary conditions is important for the finite lattice stability bounds. If e.g. periodic boundary conditions are adopted, for the free field zero mass ($m_u=0$) case, the finite lattice partition function becomes infinite due to the presence of zero eigenvalues in the quadratic form. However, in the nonzero mass case, periodic boundary conditions may be employed and the thermodynamic limit of the periodic free energy exists. Next, the zero mass limit also exists and is the same as the thermodynamic limit of the free boundary condition case, with zero mass. For these results, see \cite{M}.
\end{rema}
\begin{rema}
In the proof of Theorem \ref{thm1}, it is clear that we can also treat more general lattices $\Lambda$ verifying the condition that every site of $\Lambda$ has at least one nearest neighbor in $\Lambda$.
\end{rema}

We now turn to the pure-gauge, Yang-Mills partition function $Z^w_{Y,\Lambda,a}$ of Eq. (\ref{Zuw}). Similar scaling considerations apply here, but the treatment is more complicated as gauge invariance plays an important role. Each component of the physical gluon field is locally scaled like the massless scalar Bose field. Formally, the gluon scaling factor gives rise to the scaling factor for the unscaled Wilson partition function $Z^w_{Y,\Lambda,a}$. Using this scaling factor, the scaled pure-gauge partition function $Z_{Y,\Lambda,a}$ is defined and is expected to satisfy thermodynamic and ultraviolet stable stability bounds. Recently, in \cite{YM}, we obtained thermodynamic and ultraviolet stability bounds for the scaled Wilson lattice plaquette $Z_{Y,\Lambda,a}$. This scaled partition function is obtained by multiplying the unscaled Wilson partition function $Z^w_{Y,\Lambda ,a}$ of Eq. (\ref{Zuw}) by a scaling factor $(g^2/a^{d-4})^{-d(N)\Lambda_r/2}$.

We now give a brief description on how the scaling factor used to define the scaled partition $Z_{Y,\Lambda,a}$ appears. In section \ref{sec3}, although we do not repeat the proof of \cite{YM}, we give more details and comments on the most delicate points. For each lattice bond $b$, we parametrize the gauge bond variables $g_b$ using the physical gluon-like gauge fields
$A_b$ as ($g$ here is the gauge coupling constant appearing in the Wilson action!)
$$
g_b\,=\,e^{iagA_b}\,.
$$
For $b=b_\mu(x)$, i.e. a bond in the $\mu$-direction starting at the lattice site $x$, we identify $A_b$ with the gauge potential $A_\mu(x)$, simply denoted by $A$. Formally, in Ref. \cite{Gat}, using the Baker-Campbell-Hausdorff formula, it is shown that, for small lattice spacing $a$, the Wilson plaquette action is the Riemann sum approximation to the usual classical smooth field continuum Yang-Mills action (see Refs. \cite{Wei,Banks}), with $ \{\mu<\nu\}\,\equiv\{\mu,\nu=0,...,(d-1)\,/\,\mu<\nu\}$,
\bequ\lb{TrF2}
\mathcal A_{\mathrm{classical}}\,=\,\sum_{\{\mu<\nu\}}\,\dis\int_{{\mathbb R}^d}\;{\mathrm Tr}[F_{\mu\nu}(x)]^2\,d^dx\,=\,\sum_{\{\mu<\nu\}}\:\;\dis\int_{{\mathbb R}^d}\; {\mathrm Tr}\left\{\partial_\mu A_\nu(x)-\partial_\nu A_\mu(x)+ig[A_\mu(x),A_\nu(x)]\right\}^2\,d^dx\,.
\eequ
For each fixed $x\in\mathbb R^d$, $\mu,\nu=0,1,...,(d-1)$, we have that $\partial_\mu A_\nu(x)$, $[-\partial_\nu A_\mu(x)]$ and $i[A_\mu(x),A_\nu(x)]$ are self-adjoint, as well as their sum, which defines the antisymmetric second-order field tensor,
$$
F_{\mu\nu}(x)\,=\,\partial_\mu A_\nu(x)-\partial_\nu A_\mu(x)+ig[A_\mu(x),A_\nu(x)]\,.
$$
Hence, $[F_{\mu\nu}(x)]^2$ is self-adjoint and positive (nonnegative!), and  its trace is also positive. Cubic and quartic terms in the gauge fields are present in the non-abelian case in the continuum spacetime. The quartic term is local and positive. In the Riemann sum approximation, the derivatives become finite differences and the positivity property is maintained. The positivity property of each plaquette action $(a^{d-4}/g^2)\,\mathcal A_p$ is global and does not depend on the size of the fields. The positivity property of each plaquette action implies that Wilson plaquette action obeys positivity for the whole group, and it is also manifested for the associated Lie algebra which is written in terms of the gluon fields.

Making the parallel with the previous treatment for the Bose-gauge partition function $Z_{B,\Lambda,a}(\tilde g)$, we consider each component of $A$ as an unscaled massless scalar field. Additionally, we take the restriction to small gauge fields $A$ \ch (not needed when the abelian gauge group $\mathrm U(1)$ is considered!) and introduce the scaled fields $y$ by writing
$$
y\,=\,a^{(d-2)/2}\,A\,.
$$
Fix the enhanced temporal gauge and consider only the $\Lambda_r$ retained bonds. We transform the Wilson action, Haar measure, and hence the Wilson partition function $Z^w_{Y,\Lambda,a}$, passing to $y$ fields which yields a scaling factor
$$
\left( \dfrac{g^2}{a^{d-4}}\right)^{d(N)\Lambda_r}\,,
$$
times a partition function.

The partition function is nonsingular in $a\in(0,1]$ and $g^2$, $0<g^2\leq g_0^2$. We take it as a candidate for a scaled partition function and, based on these considerations, we define a scaled partition function $Z_{Y,\Lambda,a}$ by
\bequ\lb{ZYn}
Z_{Y,\Lambda,a}\,=\,s_Y^{d(N)\Lambda_r}\:Z^w_{Y,\Lambda,a}\quad;\quad s_Y\,=\,\dfrac{a^{(d-4)/2}}{g}\,.
\eequ

It is proved in Ref. \cite{YM} that indeed the scaled pure-gauge partition function $Z_{Y,\Lambda,a}$, for the hypercubic lattice with free boundary conditions, satisfies thermodynamic and ultraviolet stability bounds, and we state this result in Theorem 2.
\begin{thm}\lb{thm2}
Consider the scaled pure Yang-Mills, Wilson partition function $Z_{Y,\Lambda,a}$ defined in Eq. (\ref{ZYn}). For $a\in(0,1]$ and $g^2\in(0,g_0^2]$, $g_0^2<\infty$, $Z_{Y,\Lambda,a}$ satisfies the thermodynamic and ultraviolet stable stability bound
\bequ\lb{stY}
e^{c_{Y,\ell}\Lambda_r} \leq Z_{Y,\Lambda,a} \leq e^{c_{Y,u}\Lambda_r}\,,
\eequ
with finite real constants $c_{Y,\ell}$ and $c_{Y,u}$ independent of $\Lambda$, $a$ and $g^2$.
\end{thm}
\begin{rema}
As the reader may already have noticed, a key point in our method is that our lower and upper bounds on the unscaled partition functions must exhibit the same multiplicative singular factor. This is what allows us to define a scaled partition function by extracting the singular factor multiplicatively.
\end{rema}
\begin{rema}
For the special case of the abelian gauge group $\mathrm U(1)$, both the pure-gauge action and the coupling with Bose and fermi fields were treated in Refs. \cite{Driver,BalabanHiggs,King,Dim,DimockBose}. The starting point for all these works is a quadratic action for the gauge, electromagnetic potentials (fields). In this approach, to remove the null space of the quadratic form and define the starting partition function, a gauge fixing is required at the onset. This is not what it is done here! A rigorous connection with the Wilson partition function with the Wilson plaquette action, used here and which is {\em not} quadratic, has not been established.
\end{rema}
\begin{rema}
For the special case of the gauge group $\mathcal G\,=\,{\mathrm SU}(2)$,
there is a $1-1$ correspondence between the group and the unit sphere $S^3$ in $\mathbb R^4$  (see Refs. \cite{Simon2,Hall}). In terms of gluon fields, an explicit formula for the Haar measure is available and a direct proof of our thermodynamic and ultraviolet stable stability bounds is given in Appendix B. For the gauge groups ${\mathrm SU}(N\geq 3)$, explicit expressions for the Haar measures also exist \cite{Nelson,Holland,Marinov,Euler}, using several group element parametrizations, but either it is not evident how to determine an integration domain of parameters such that the parametrization is $1-1$ and onto the group or the parametrization does not lend itself to derive a quadratic bound on the action.
\end{rema}
\begin{rema}
It is expected that the control over global quantities, such as thermodynamic and ultraviole stability bounds for the partition function, lead to control over local quantities such as generating functions for correlations. This expectation is confirmed in Ref. \cite{GYM}, where we proved boundedness, uniformly in $a\in(0,1]$, of the infinite lattice $a\mathbb Z^d$ normalized $r$-plaquette field generating function. (The definition of the plaquette field generating function is inspired by the work of Ref. \cite{Schor}.) After extending the stabilty bound to this case, the well known method of multiple reflections (see \cite{GJ}) is used for our model with periodic boundary conditions. These results hold whether or not the mass gap persists as $a\searrow 0$. 
\end{rema}

From Theorems \ref{thm1} and \ref{thm2}, recalling that $\Lambda_r\simeq (d-1)\Lambda_s$ (for $L\gg 1$), and applying the generic bound of Eq. (\ref{principal}) to the complete scaled partition function
\bequ\lb{nZ}
Z_{\Lambda,a} \, \equiv \, s_B^{N\Lambda_s}\,s_Y^{d(N)\Lambda_r}\,Z^u_{\Lambda,a}\,,
\eequ
with $Z^u_{\Lambda,a}$ given in Eq. (\ref{Z}), we obtain the stability bound given in the next theorem, for our complete Bose-gauge model.
\begin{thm}	\lb{thm3}
The scaled partition function $Z_{\Lambda,a}$ of Eq. (\ref{nZ}) satisfies the thermodynamic and ultraviolet stable stability bounds
\bequ\lb{stZ}
e^{c_{\ell}\Lambda_s} \leq Z_{\Lambda,a} \leq e^{c_{u}\Lambda_s}\,,
\eequ
with finite real constants $c_{\ell}=c_{B,\ell}+c_{Y,\ell}\Lambda_r/\Lambda_s$ and $c_{u}=c_{B,u}+c_{Y,u}\Lambda_r/\Lambda_s$ independent of the size $L$ of $\Lambda$, $a$ and $g^2$.
\end{thm}
\begin{rema}
For the special case of the abelian gauge group $\mathrm U(1)$, gauge-matter models with either Bose or fermi fields were treated for dimensions $d=2,3$ in Refs. \cite{BalabanHiggs,King,DimockBose,Dim}. Again, in these works, the pure gauge action is quadratic from the beginning and gauge fixing is necessary to define a finite-lattice partition function.
\end{rema}

Applying the Bolzano-Weierstrass theorem, from Theorem \ref{thm3}, we have the existence of the normalized free energy $f^n_{\Lambda,a}$. This is the content of the following corollary.
\begin{corollary}\lb{coro1}
The normalized free energy per lattice site defined in Eq. (\ref{fn}) satisfies the bounds
\bequ\lb{nFE}c_\ell\,\leq\,f^n_{\Lambda,a}\,\leq\, c_u\,,\eequ
so that $f^n_{\Lambda,a}$ has, at least in the sense of subsequences, a finite thermodynamic limit
$f^n_{a}$, when $\Lambda\nearrow a\mathbb Z^d$. Subsequently, at least, a finite subsequential continuum limit free energy $f^n\,\equiv \lim_{a\searrow 0}f^n_a$ also exists.
\end{corollary}
\begin{rema}
We point out that our result is focused only on the existence of the limiting free energy. We do not make any claim about the spectral properties of the corresponding models, such as the existence of particles, their masses and scattering.
\end{rema}

Before discussing the proofs of the above theorems, we make some comments on the structure of the upper and lower stability bounds for the partial scaled partition functions $Z_{B,\Lambda,a}(\tilde g)$ and $Z_{Y,\Lambda,a}$.

The upper bound for $Z_{B,\Lambda,a}(\tilde g)$ is reduced to a product over bonds of a bound on a single-bond {\em transfer matrix}; the single-bond gauge variable can be taken to be equal to the identity gauge group matrix, by gauge invariance. The lower bound on $Z_{B,\Lambda,a}(\tilde g)$ follows directly from Jensen's inequality.

For the upper bound on $Z_{Y,\Lambda,a}$ of Eq. (\ref{ZYn}), we do not  use the enhanced temporal gauge. Instead, the upper bound on $Z_{Y,\Lambda,a}$ follows from an upper bound on $Z^w_{Y,\Lambda,a}$ which uses, as a main ingredient, the pointwise positivity of each plaquette action $(a^{d-4}/g^2)\mathcal A_p=(a^{d-4}/g^2)\|1-g_p\|^2_{H-S}$ of Eq. (\ref{YMaction}). The upper bound is reduced to a product over retained bonds of a bound on a single plaquette, single-bond variable partition function $z$. Even though we have not imposed any gauge fixing, such as the enhanced temporal gauge, the number of factors is the number of retained bonds $\Lambda_r$. The lower bound on $Z^w_{Y,\Lambda,a}$ is also a product over retained bonds of a single plaquette, single bond partition function $\tilde z$. There are no small field restrictions imposed to obtain the factorized bound. However, to obtain a lower bound on the single-bond variable, single plaquette partition function $\tilde z$, and to extract a factor $(g^2/a^{d-4})^{d(N)/2}$, we impose a small field condition.  Besides, using the positivity of the integrand, we have the luxury of restricting the bond variables and requiring the group elements to be near the identity element. For the lower bound, a new upper global bound for the plaquette action is proved, which is quadratic in the plaquette gluon fields. That this bound is quadratic is surprising as the naive small $a$ approximation has positive quartic terms in the non-abelian case.

We remark that the lower bound on $Z^w_{Y,\Lambda,a}$ follows from a global {\em quadratic} pointwise upper bound on the plaquette action. To prove the upper bound on $Z^w_{Y,\Lambda,a}$, we need a lower bound on the action. But, as it can already be seen explicitly when dealing with the abelian case $\mathcal G\,=\,\mathrm U(1)$ (see below), there is no global quadratic lower bound for the single plaquette action. However, it turns out that we do have a global lower bound for the single-bond single-plaquette action in $z$. This is so because, in $z$, we are left with only a single bond variable.

Both the upper and the lower stability bounds on  $Z^w_{Y,\Lambda,a}$ are proven in Ref. \cite{YM}. Here, we review the proofs. More than this, we give a new proof for the lower bound which gives a tremendous simplification of the proof given in \cite{YM}.

We now turn to $d=2$, where we make a connection between the Circular Unitary Ensemble (CUE) and the Gaussian Unitary Ensemble (GUE) of random matrix theory (see \cite{Metha,Deift,AGZ}). We also obtain an explicit formula for the scaled free energy in the thermodynamic and continuum limits. For $d=2$, an exact result is obtained for $Z_{Y,\Lambda,a}$ in \cite{Ash} using a character expansion. Our upper bound holds for $d=2,3,4$ and reads
$$
Z^w_{Y,\Lambda,a}\,\leq\,z^{\Lambda_r}\,,
$$
where, with an integral over a single bond group variable $U$,
\bequ\lb{z}
z\,\equiv z(a)\,=\,\dis\int\,\exp\left[-\dfrac{a^{d-4}}{g^2}\,A(U)\right]\,d\sigma(U)\,,
\eequ
and with $A(U)={\mathrm Tr}[2-U-U^{-1}]$. The upper bound is exact for $d=2$ and $\Lambda_p=\Lambda_r=(L-1)^2$.

Defining the unnormalized pure-gauge free energy per retained bond by
$$f_{Y,\Lambda,a}\,=\,\dfrac1{\Lambda_r}\,\ln Z^w_{Y,\Lambda,a}\,,$$
we see that the thermodynamic limit of $f_{Y,\Lambda,a}$ is $\ln z$.

In the sequel, an important concept is that of a class or central function on the group $\mathcal G$. For $g\in\mathcal G$, if a function $f(g)$ is constant in the conjugacy classes of $\mathcal G$, i.e. if $f(g)$ satisfies $f(g)\,=\,f(hgh^{-1})$, for all $h\in\mathcal G$, we say that $f$ is a class function. For a class function on $\mathcal G$, the integral over the group can be expressed as an integral over the angular eigenvalues of the group element. (Note that this is also a global property which does not depend on a restriction on the group elements!). Using the Weyl integration formula for a class function on $\mathcal G\,=\,{\mathrm U}(N),\,{\mathrm SU}(N)$ (see Refs. \cite{Weyl,Metha,Simon2,Bump,Deift}), the single-bond partition function $z$ has a representation in terms of an integral over the distribution of angular eigenvalues of a unitary matrix in CUE.

Specializing to the case of the gauge group $\mathrm U(N)$, where $d(N)\,=\,N^2$, by the spectral theorem, a unitary matrix $U$ has eigenvalues $e^{i\la_1}$, ..., $e^{i\la_N}$, with the angular eigenvalues $\la=(\la_1,\ldots,\la_N)\in(-\pi,\pi]^N$. Then,
$$f_{Y,\Lambda,a}\,=\,\ln z\,,
$$
with
\bequ\lb{z} z\,=\,\dfrac1{\mathcal N_C(N)}\,\dis\int_{(-\pi,\pi]^N}\,\exp\left[-\dfrac{a^{d-4}}{g^2}\,\sum_{j=1}^N\,2(1-\cos \la_j)\right]\,\rho(\la)\,d^N\la\,,
\eequ
where $\mathcal N_C(N)\,=\,(2\pi)^NN!$ and $[\rho(\la)\,d^N\la/\mathcal N_C(N)]$ is the Haar eigenvalue measure. This probability distribution is also referred to as the CUE. Here, the angular eigenvalue density is related to a Vandermonde determinant and reads \bequ\lb{dense}\rho(\la)\,=\,\prod_{1\leq j<i\leq N}\,|e^{i\la_j}-e^{i\la_i}|^2\,=\,
\prod_{1\leq j<i\leq N}\,\{2[1-\cos(\la_j-\la_i)]\}\,.\eequ

From this representation, we can determine $f_{Y}$, the thermodynamic and the continuum limit of $f_{Y,\Lambda,a}$, as an integral over the probability distribution of eigenvalues of a self-adjoint matrix in the GUE. The GUE has the probability measure in $\mathbb R^n$, with a normalization $\mathcal N_G(N)\,\equiv\,I(\infty)\,=\,(2\pi)^{N/2}\,2^{-N^2/2}\prod_{1\leq j\leq N}j!$ (see Ref. \cite{Metha}),
$$\dfrac1{\mathcal N_G(N)}\,
e^{-y^2}\,\hat\rho(y)\,d^Ny\qquad,\qquad \hat\rho(y)\,=\,\prod_{1\leq j<k\leq N}(y_j-y_k)^2\,.
$$
Below, we need the integral of this distribution over the bounded domain $(-u,u]^N$, and we denote it by
\bequ\lb{I}
I(u)\,\equiv\,\dis\int_{(-u,u]^N}\,e^{-y^2}\,\hat\rho(y)\,d^Ny\,\leq I(\infty)\,=\,\mathcal N_G(N)\,=(2\pi)^{N/2}\,2^{-N^2/2}\prod_{1\leq j\leq N}j!\,,
\eequ
where the inequality follows from the fact that $I(u)$ is bounded and monotone increasing, and satisfies $I(u)\leq I(\infty)=(2\pi)^{-N/2} \prod_{j=1}^{N-1}\,j!$

Thus, the GUE distribution appears in a natural way, and we give this interesting connection as a corollary to the following theorem that connects the limit of the CUE with the GUE.
\begin{thm}\lb{thm4}
Let $w$ be given by
\bequ\lb{w4}
w\,=\,\dfrac1{\mathcal N_C(N)}\,\dis\int_{(-\pi,\pi]^N}\,e^{-L(\la)/\beta}\,\rho(\la)\,d^N\la\,,
\eequ
where $\beta>0$ and $\rho(\la)\,=\,\prod_{1\leq j<k\leq N}\,|e^{i\la_j}-e^{i\la_k}|^2$ and $\rho(\la)d^N\la/(N!)$ is the normalized angular eigenvalue Haar measure for $\mathcal G\,=\,{\mathrm U}(N)$. If, for some constant $c$, verifying $0<c<\infty$, the function $L(\la)$ obeys the lower bound $L(\la)\,\geq\,c\la^2$, $\la\in(-\pi,\pi]^N$, and $\lim_{\beta\rightarrow 0}\,\left[L(\sqrt\beta y)/\beta\right]\,=\,y^2$, for all $y\in(-\pi/\sqrt\beta,\pi/\sqrt\beta]^N$, then, recalling that $d(N)\,=\,N^2$, for $\mathcal G=\mathrm U(N)$,
$$\barr{lll}
\lim_{\beta\rightarrow 0}\dfrac{w}{\beta^{N^2/2}}&=&\dfrac1{\mathcal N_C(N)}\,\dis\int_{\mathbb R^N}\,e^{-y^2}
\prod_{1\leq j<k\leq N}\,(y_j-y_k)^2\,d^Ny\,=\,\dfrac{\mathcal N_G(N)}{\mathcal N_C(N)}\,=\,2^{-[N^2-N]/2}\,\pi^{-N/2}\,\prod_{1\leq j\leq N-1}j!\,.\earr$$
The integral is precisely the normalization for the distribution of the GUE.
\end{thm}
\begin{rema}In Ref. \cite{YM}, it is shown that the same type of result holds also for $\mathcal G\,=\,{\mathrm SU}(N)$.
\end{rema}

We apply Theorem \ref{thm4} to obtain the continuum limit $a\searrow 0$ of the normalized pure-gauge free energy in $d=2$. The normalized free energy per retained bond is defined by
\bequ\lb{fYn}f^n_{Y,\Lambda,a}\,=\,\dfrac1{\Lambda_r}\,\ln Z_{Y,\Lambda,a}\,,\eequ
where $Z_{Y,\Lambda,a}\,=\,(ga)^{N^2\Lambda_r}\,z^{\Lambda_r}$, so that $f^n_{Y,\Lambda,a}\,=\,\ln z\,-\,N^2\ln(ga)$.

In Theorem \ref{thm4}, make the  identifications $\beta\,=\,(ga)^2$, $z\,=\,w$ and $L(\la)\,=\,2\sum_{j=1,\ldots,N}(1-\cos\la_j)$. Using the lower bound (see \cite{Simon3}), $1-\cos u\geq 2u^2/\pi^2$, $|u|\leq \pi$, we have $L(\la)\,\geq\,4\la^2/\pi^2$, so that we can take $c=4/\pi^2$ in Theorem \ref{thm4}. Furthermore, $$\lim_{\beta\searrow 0}\left[ L(\sqrt\beta y)/\beta\right]\,=\,\lim_{\beta\searrow 0}\left[\dfrac 2\beta\,\sum_j\,\left(1-\cos \sqrt\beta y_j\right)\right]\,=\,y^2\,.$$ Hence, by Theorem \ref{thm4}, as $\beta\searrow 0$,
$$
\dfrac z{\beta^{d(N)/2}}\,=\,\dfrac z{(ga)^{N^2}}\,\rightarrow\,\dfrac{\mathcal N_G(N)}{\mathcal N_C(N)}\,.
$$

Finally, taking the logarithm, as a corollary to Theorem \ref{thm4}, we can state:
\begin{corollary}\lb{coro2}
The continuum limit of the normalized free energy $f^n_{Y,\Lambda,a}$ of the pure gauge model in $d=2$ exists, is finite, and is given by
$$
f_{Y}^n\,\equiv\,\lim_{a\searrow 0}\;\lim_{\Lambda\nearrow a\mathbb Z^d}\,f^n_{Y,\Lambda,a}\,=\,\ln\left( \dfrac{\mathcal N_G(N)}{\mathcal N_C(N)}\right)\,=\,\ln\left(2^{-[N^2-N]/2}\,\pi^{-N/2}\,\prod_{1\leq j\leq N-1}j!\right)\,.
$$
\end{corollary}

Returning to the $d=2,3,4$ cases, an important remark is that thermodynamic and ultraviolet stable stability bounds can be obtained in a much easier way when the gauge group is ${\mathrm SU}(2)$. In this special case, we use the gluon fields and show that this gives the same result as the Weyl angular eigenvalue integration formula. This is so because ${\mathrm SU}(2)$ can be identified with the unit sphere $S^3$ in $\mathbb R^4$ and we can determine the minimal domain for which there is a $1-1$ map from the group elements and the gluon fields in the associated Lie algebra so that we can easily control the gauge integrals. The simple case of the gauge group $\mathcal G\,=\,{\mathrm SU}(2)$ is treated in Appendix B.

In closing this section, we mention that in Refs. \cite{MP,MP2} we have also considered gauge models with Fermi matter fields, like the quarks in traditional QCD. By using a vertex expansion and spectral representations, we are able to prove an upper stability bound. However, up to now, we do not know of any substitute method to Jensen's inequality used here to obtain lower stability bound, whenever fermions are present. Together with confinement, this is a step to overcome if we want to show the existence of traditional QCD.

The rest of the paper is organized as follows. In section \ref{sec2}, we prove Theorem \ref{thm1} for real Bose fields. The proof bypasses diamagnetic inequality considerations. The proof of Theorem \ref{thm2} is given in Ref. \cite{YM}. In section \ref{sec3}, we outline and greatly improve the proof of Theorem \ref{thm2}. We obtain a new global quadratic upper bound on the Wilson plaquette action $(a^{d-4}/g^2)\mathcal A_p$, which is quadratic in the fields. In the end of section \ref{sec3}, we prove Theorem \ref{thm4}. Section \ref{sec4} is devoted to some concluding remarks and in Appendix A, we extend the proof of Theorem \ref{thm1} to complex Bose fields. In Appendix B, we first give some well known facts about ${\mathrm SU}(2)$ and a new result on the inverse of the exponential map from the Lie algebra to the group. Next, by exploring the simple relation between the gluon fields and angular eigenvalues, we show stability bounds for the special case of the gauge group $\mathcal G\,=\,{\mathrm SU}(2)$.
%%%%%%%%%%%%%%%%%%%%%%%%%%%%%%%%%%%%%%%%%%%%%%%%%%%%%%%%%%%%%%%%%%%%%%%%%%
%%%%%%%%%%%%%%%%%%%%%%%%%%%%%%%%%%%%%%%%%%%%%%%%%%%%%%%%%%%%%%%%%%%%%%%%%%
\section{Stability Bounds on the Bose-Gauge Scaled Partition Function $Z_{B,\Lambda,a}(\tilde g)$} \lb{sec2}
%%%%%%%%%%%%%%%%%%%%%%%%%%%%%%%%%%%%%%%%%%%%%%%%%%%%%%%%%%%%%%%%%%%%%%%%%%
%%%%%%%%%%%%%%%%%%%%%%%%%%%%%%%%%%%%%%%%%%%%%%%%%%%%%%%%%%%%%%%%%%%%%%%%%%
Here, for simplicity, we explicitly prove Theorem \ref{thm1} for the scaled real scalar Bose field. It suffices to consider the orthogonal gauge group $\mathcal G={\mathrm O}(N)$, instead of $\mathcal G={\mathrm U}(N)$. The new proof bypasses diamagnetic inequality considerations employed in \cite{BFS} and uses the H\"older's inequality to bound the partition function by a product of one-dimensional chain partition functions.
Although the gauge field in $Z_{B,\Lambda,a}(\tilde g)$ is arbitrary, the chain partition function $z$ is independent of the gauge field. The gauge variable can be set to the identity for all bonds in the chain. This bound is like the case of the diamagnetic inequality in the sense that our bound is independent of the gauge field. The use of free boundary conditions is important. For periodic boundary conditions, there is a dependence on the gauge fields. The chain partition function can be made to depend on only one bond of the chain, where the gauge field is a product of all gauge fields in the chain.

For free boundary conditions, the bound on each chain partition function is reduced to a product of bounds of a single-bond 'transfer matrix'. The extension to complex Bose fields is succinctly given in Appendix A.

For the reader's convenience, we first give some elementary inequalities that we use. For the integral of a product of complex valued functions, we have H\"older's inequality for complex valued functions (see \cite{RS1}),
\bequ\lb{Holder}
\int\,|f_1(x)\dots f_n(x)|\,d\hat\mu(x)\,\leq\,\prod_{j=1,\ldots ,n}\,\left[\int\,|f_j(x)|^{p_j}\,d\hat\mu(x)\right]^{1/p_j}\quad;\quad \sum_{j=1}^n\, \dfrac 1{p_j}\,=\,1\,,
\eequ
with $1\leq p_j<\infty$ and $d\hat\mu$ is a measure.
Jensen's inequality, for real valued $f(x)$, is used to obtain lower bounds and is given by (see e.g. \cite{Rudin})
\bequ\lb{Jensen}
\int\,e^{f(x)}\,d\hat\mu(x)\,\geq\,\exp\left[{\int\, f(x)\,d\hat\mu(x)}\right]\qquad;\qquad {\mathrm if\;\,}\int \,d\hat\mu(x)\,=\,1\,.
\eequ
Also, for a generic linear operator $T$ in an Euclidean or complex vector space with an inner product $(\cdot,\cdot)$ and vector norm $|f|$, we have the vector space induced operator norm or, simply, operator norm \cite{RS1}
\bequ\lb{Holmgren}
\| T\|\,=\,\sup_{f:\,|f|=1}\,|Tf|\,=\,\sup_{f,g:\, |f|=1,\,|g|=1}\,|(g,Tf)|\,,
\eequ
and for the operator norm, we have Holmgren's inequality (see Ref. \cite{Simon3})
\bequ\lb{Holmgren1}
\| K\|\,\leq \left[ \sup_x\,\dis\int |K(x,y)|\,dy \right]^{1/2}\;\left[ \sup_y\,\dis\int |K(x,y)|\,dx \right]^{1/2}\,,
\eequ
where $K(x,y)$ is the kernel of an integral operator $K$ acting on the space of square integrable functions. Namely, we have $[Kf](x)\,=\,\int\,K(x,y)f(y)\,dy$. In terms of the kernel $K$, the Hilbert-Schmidt norm, denoted by $\|\cdot\|_{H-S}$, is $\left[\int |K(x,y)|^2 dxdy\right]^{1/2}$. We have $\|K\|\,\leq\,\|K\|_{H-S}$.

The Hilbert-Schmidt norm for a $n\times n$ matrix $A\,=\,\left(  a_{ij}\right)$, is defined by $$\left\|A\right\|_{H-S}\,\equiv\,\left[Tr(A^\dagger\,A)\right]^{1/2}\,.$$
Taking another $n\times n$ matrix $B\,=\,\left(  b_{ij}\right)$, we also define the inner product $(A,B)\,=\,Tr(A^\dagger\,B)$ and use $\|\,.\,\|$ to denote the usual operator (matrix) norm. Both norms satisfy the submultiplicative property
$$
\left\|AB\right\|_{H-S}\,\leq\,\left\|A\right\|_{H-S}\,\left\|B\right\|_{H-S}\qquad;\qquad \left\|AB\right\|\,\leq\,\left\|A\right\|\,\left\|B\right\|\,.
$$

We now turn to the proof of Theorem \ref{thm1} for the real Bose field case, with one flavor ($N_f=1$), and the real orthogonal group $\mathcal G={\mathrm O}(N)$. The complex extension of this case is given in Appendix A. The extension to multiflavored fields and to more general random matrix  fields is direct.

For real Bose fields, the two terms in the hopping action of Eq. (\ref{sM}) are equal. Thus, we maintain only the first one and let $\kappa^2\rightarrow 2\kappa^2$. The lower stability bound $Z_{B,\Lambda,a}(\tilde g)\,\geq\,1$ follows from the application of Jensen's inequality of Eq. (\ref{Jensen}) to the integral
$$
Z_{B,\Lambda,a}(\tilde g)\,=\,\dis\int\,\prod_{b_\mu(x)}\,F_{b_\mu(x)}\,d\nu(\tilde\phi)\,.
$$
The measure is a normalized Gaussian probability measure $d\nu({\tilde\phi})\,=\,\prod_x (2\pi)^{-1/2}\,e^{-\phi^2(x)/2}\,d\phi(x)$.

For the upper bound, we use H\"older's inequality to factor over the $d$ spacetime directions when considering   $\prod_{\mu=0,...,d-1}\,\exp[\kappa^2\sum_{x}\phi(x)g_{b_\mu(x)}\phi(x^+_\mu)]$ in the integrand. Doing this in Eq. (\ref{ZM}), we obtain
$$
Z_B\,\equiv Z_{B,\Lambda,a}(\tilde g)\,\leq\,\prod_{\mu=0}^{d-1}\,\left[ \dis\int\, e^{d\kappa^2\sum_{x}\phi(x)g_{b_\mu(x)}\phi(x^+_\mu)}\,d\nu(\tilde\phi)\right]^{1/d}\,,
$$
Note the factor $d$ in the exponent of the above integral.

Next, each integral factorizes into a product of partition functions, each of which is a partition function of a one-dimensional chain. A generic connected chain partition function $Z_c$, with $L$ sites and $L-1$ bonds has the representation
\bequ\lb{Zcc}
Z_c\,=\,\dfrac 1 {(2\pi)^{LN/2}}\,\dis\int\,h(x_1)\,\prod_{j=1}^{L-1}\;\left[T_b(x_{j+1})\right]\,h(x_L)\,dx_1\ldots dx_L\,,
\eequ
where $h(x)\,=\,e^{-x^2/4}$.
The free-boundary chain partition function $Z_c$ is actually independent of the bond variables $g$ of the chain. This can be seen by performing successive change of variables along the chain to absorb the $g$ factors. The kernel $T_b(x,y)$ is given by
\bequ\lb{kerT}
T_b(x,y)\,=\,e^{-x^2/4}\,e^{d\kappa^2xy}\,e^{-y^2/4}\,,\eequ
and has the limit $e^{(x-y)^2/4}$ as $a\searrow 0$. Due to the translation invariance of $e^{(x-y)^2/4}$, the Hilbert-Schmidt norm is infinite.
The integral on the right-hand-side of Eq. (\ref{Zcc}) is an inner product in the space $L^2(\mathbb R^N)$, and we have [see Eq. (\ref{kerT})]
$$
Z_c\,=\,\dfrac 1 {(2\pi)^{LN/2}}\;\left(h,\,\prod_b [T_b(g_b)]h\right)\,\leq\,\dfrac 1 {(2\pi)^{LN/2}}\,|h|^2\;\prod_b\,\| T_b(g_b)\|\,\leq\,\dfrac 1 {(2\pi)^{LN/2}}\,|h|^2\;\prod_b\,\| T_b(g_b)\|^{L-1}\,,
$$
where we recall that $\|\cdot\|$ denotes the operator norm in the space $L^2(\mathbb R^N)$.

Now, we apply the Holmgren bound of Eq. (\ref{Holmgren1}) to $\| T_b(g_b)\|$ which gives
$$
\| T_b(g_b)\|\,\leq\,\left[ \sup_{\phi_1}\,e^{-\phi_1^2/4}\dis\int\, e^{d\kappa^2(\phi_1,g_b\phi_2)}e^{-\phi_2^2/4}\,d\phi_2\right]^{1/2}
\;\left[ \sup_{\phi_2}\,e^{-\phi_2^2/4}\dis\int\, e^{d\kappa^2(\phi_1,g_b\phi_2)}e^{-\phi_1^2/4}\,d\phi_1\right]^{1/2}\,.$$
Recalling Eq. (\ref{hopbose}) and calculating the first integral yields $(4\pi)^{N/2}\,e^{d^2\kappa^4\phi_1^2}\,\leq\,(4\pi)^{N/2}\,e^{\phi_1^2/4}$. The same result holds for the other integral with $\phi_1\rightarrow \phi_2$. Thus,
$$
\| T_b(g_b)\|\,\leq\,(4\pi)^{N/2}\,,
$$
and $Z_c\,\leq\,(4\pi)^{N(L-1)/2}$. Using this result in $Z_B\,\equiv\,Z_{B,\Lambda,a}(\tilde g)$, we obtain
$$
Z_B\,\leq\,2^{N(L-1)L^{d-1}/2}\,\leq\,e^{c_{B,u}\Lambda_s}\,,
$$
where $c_{B,u}\,\geq\,N(1-L^{-1})\ln 2/2$.

If the chain is not connected, for example, a thick `U' shaped region has disconnected chains, the above analysis still applies for each connected component. Also, we remark that using the Holmgren estimate is crucial for this upper bound; the Hilbert-Schmidt norm of $T_b$ diverges when $a\searrow 0$. This ends the proof of Theorem \ref{thm1} for the real Bose case. The extension to the complex Bose field case is given in Appendix A.\qed
%%%%%%%%%%%%%%%%%%%%%%%%%%%%%%%%%%%%%%%%%%%%%%%%%%%%%%%%%%%%%%%%%%%%%%%%%%
%%%%%%%%%%%%%%%%%%%%%%%%%%%%%%%%%%%%%%%%%%%%%%%%%%%%%%%%%%%%%%%%%%%%%%%%%%
\section{On the Stability Bounds for the Scaled Yang-Mills Partition Function $Z_{Y,\Lambda,a}$} \lb{sec3}
%%%%%%%%%%%%%%%%%%%%%%%%%%%%%%%%%%%%%%%%%%%%%%%%%%%%%%%%%%%%%%%%%%%%%%%%%%
%%%%%%%%%%%%%%%%%%%%%%%%%%%%%%%%%%%%%%%%%%%%%%%%%%%%%%%%%%%%%%%%%%%%%%%%%%
The complete proof of Theorem \ref{thm2} appears in \cite{YM}. In subsection III.1 we give a detailed derivation leading to the choice of the scaling factor of Eq. (\ref{ZYn}) used to define the scaled pure Yang-Mills partition function $Z_{Y,\Lambda,a}$ of Eq. (\ref{ZYn}). In subsection III.2, we give the  main ingredients and steps in the proof of Theorem \ref{thm2}. The gauge is not fixed in the proof of the upper stability bound. The enhanced temporal gauge fixing is used only to prove the lower stability bound. Also, to obtain the lower bound, we prove a new upper global quadratic bound in the gluon fields to the single-plaquette four-bond Wilson action $(a^{d-4}/g^2)\mathcal A_p$. This upper bound is a class function in each bond variable. This new result improves and simplifies enormously the bound given in Lemma 1 of \cite{YM}, where the gluon fields were restricted to be small. \ch (This restriction is not needed if the gauge group is the abelian group $\mathrm U(1)$). We close this subsection with the proof of Theorem \ref{thm4}. Both the upper and lower bounds factorize into the product of a single plaquette partition function of a single-bond variable. The number of factors is $\Lambda_r$.

Concerning the proof of Theorem \ref{thm2}, first we give intuition for the choice of the scaling factor of Eq. (\ref{ZYn}) for the unscaled Yang-Mills partition function $Z^w_{Y,\Lambda,a}$. It is the unscaled partition function $Z_{Y,\Lambda,a}$ that satisfies thermodynamic and ultraviolet stability bounds with the same singular factor in the upper and lower bound. Our analysis follows the same scheme we used in dealing with the Bose-gauge partition function. The main difference here is that, to talk about the gauge potentials related to each lattice bond $b$ (the physical gluons $A_b$) which are in the Lie algebra associated to $\mathcal G$,  we need to consider gauge-group elements in $\mathcal G$ near the identity element. However, we obtain the factorized lower bound without imposing any restrictions on the bond variables. We find that we can treat each component of the gluon field in a similar manner as the massless real scalar Bose field of sections \ref{sec1} and \ref{sec2}. We call the scaled gauge fields $y$ and, in terms of these fields, the Wilson action is not singular in the lattice spacing $a$ and the gauge coupling $g$ parameters, and the scaling factor for the Wilson partition function emerges. This scaling factor is used to defined the scaled partition function $Z_{Y,\Lambda,a}$; the candidate we expect to satisfy the thermodynamic and ultraviolet stability bounds of Theorem \ref{thm2}. Because of the scaling, these bounds show the same singularity structure in the free energy when $\Lambda\nearrow a\mathbb Z^d$ and $a\searrow 0$ in the upper and lower bound, leading to the existence of the free energy per degree of freedom.

Considering  the upper bound on the scaled partition function $Z_{Y,\Lambda,a}$, since the Wilson plaquette action is pointwise positive, we have to consider a global bound on $\mathcal G$. For the lower bound, again using the positivity of the Wilson action, we have the luxury of restricting the bond variables and taking only group elements near the identity.

In the next subsections, we determine the scaling factor used to define the scaled pure-gauge partition function and deal with the upper and lower stability  bounds. We also state and prove the new global upper quadratic bound on the four-bond plaquette action $(a^{d-4}/g^2)\mathcal A_p$ and end by giving the proof of Theorem \ref{thm4}.
%=============================================================
%=============================================================
%=============================================================
\subsection{Scaling Factor for the Wilson Partition Function}
%==============================================================================%====================================================
%=====================================================
Here, after fixing the enhanced temporal gauge, we give the intuition behind the choice of the multiplicative scaling factor used to define the scaled Yang-Mills partition function $Z_{Y,\Lambda,a}$. We formally show that, for all retained bond variables near the identity, if each component of the physical gauge potentials, or gluon fields, is locally scaled by the factor $a^{(d-2)/2}$, similar to the free massless scalar Bose fields done in the previous section, then the resulting action is not singular for small $a\in(0,1]$ and $g_0\in(0,g_0]$, $0<g_0<\infty$.
The field scaling gives rise to a scaling factor for the original Wilson partition function $Z^w_{Y, \Lambda,a}$, and a scaled partition function $Z_{Y,\Lambda,a}$ which satisfies the thermodynamic and ultraviolet stable stability bounds.

Here, we write the partition function $Z^w_{Y, \Lambda,a}$ as
$$
Z^w_{Y, \Lambda,a}\,=\,\dis\int\,\exp\left\{ -\dfrac{a^{d-4}}{g^2}\sum_p \mathcal A_p(g_p)\right\}\,\prod_{b}d\mu(g_b)\,,
$$
and explain in detail how the scaling factor comes about from this expression.

For all group variables $g_b$, the physical gluon fields appear as Lie algebra parameters through the relation ($g$ here is the gauge coupling!)
$g_b=e^{igaA_b}$, with $A_b=\sum_{1\leq\alpha\leq d(N)}\,A_b^\alpha\theta_\alpha$, $\theta_\alpha\in\mathbb R$, so that $g_b\,=\,e^{iX_b}$, $X_b\,=\,\sum_\alpha\,x_b^\alpha\theta_\alpha$. If $b=b_\mu(x)$, we identify $A_b$ with $A_\mu(x)$,
the $N\times N$ matrix valued physical gauge potentials or gluon fields.

Formally, in \cite{Gat}, it is shown that the Wilson action $(a^{d-4}/g^2)\sum_p \mathcal A_p$ is, for small $a$, the Riemann sum approximation to the smooth field continuum Yang-Mills action of Eq. (\ref{TrF2}). In $Z^w_{Y,\Lambda,a}$, if, instead of using the physical gluon potential parametrization, we parametrize the bond variable $g_b\in\mathcal G$ by the fields $X_b$, so that $g_b=e^{iX_b}$, then the corresponding Haar measure is approximately a constant $c$ times the Lebesgue measure $d^{d(N)}x_b$ and
$$
Z^w_{Y,\Lambda,a}\,\simeq\,\dis\int\,\exp\left\{ -\dfrac{a^{d-4}}{g^2}\sum_p \mathcal A_p(g_b=e^{iX_b};b\in p)\right\}\,\prod_{b}c\;d^{d(N)}x_b\,.
$$

Now, the $X_b$ fields are related to the physical gluon fields $A_b$ by the local scaling relation
$$
agA_b\,=\,X_b\,.
$$
We denote by $y_b$ the scaled fields which are related to $A_b$ by the local scaling relation
$$y_b\,=\,a^{(d-2)/2}\,A_b\,.$$
For small $a$, the $a$ dependence of this relation is the same as the relation between the scaled massless scalar Bose field and its corresponding unscaled field (see section \ref{sec1}). Namely, each component of $A_b$ is taken to correspond to an unscaled massless scalar Bose field and each component of $y_b$ is the corresponding scaled massless Bose field.

By the above, the scaled gauge field $y_b$ is related to the $X_b$ field by
$$
y_b\,=\,a^{(d-2)/2}\,A_b\,=\,a^{(d-2)/2}\dfrac1{ag}X_b\,=\,\dfrac{a^{(d-4)/2}}{g}\,X_b\,.
$$
Finally, in terms of the $y_b$ fields, the partition function $Z^w_{Y,\Lambda,a}$ becomes
$$
Z^w_{Y,\Lambda,a}\,\simeq\,\left[ga^{(4-d)/2}\right]^{d(N)\Lambda_r}\,
\dis\int\,\exp\left\{-\dfrac{a^{d-4}}{g^2}\sum_p \mathcal A_p(g_b=e^{i\frac{g}{a^{(d-4)/2}}y_b\theta})\right\}\,\prod_{b}c\,d^{d(N)}y_b\,.
$$
The exponent of the exponential in the integrand is independent of $a$ and $g^2\leq g_0^2$, $0<g_0<\infty$.

Based on this approximate small field relation, we define a scaled partition function $Z_{Y,\Lambda,a}$ by
$$
Z_{Y,\Lambda,a}\,=\,\left[g^2a^{4-d}\right]^{-d(N)\Lambda_r/2}\,Z^w_{Y,\Lambda,a}\,.
$$
It is precisely this partition function which is proved to obey thermodynamic and ultraviolet stable stability bounds in Ref. \cite{YM}, as stated in Theorem \ref{thm2} above.
%=============================================================
%=============================================================
\subsection{Upper and Lower Stability Bounds on the Scaled Yang-Mills Partition Function $Z_{Y,\Lambda,a}$}
%=============================================================
%=============================================================
To obtain the upper and lower stability bounds on the scaled pure-gauge partition function $Z_{Y,\Lambda,a}$ given in Theorem \ref{thm2},
we start by analyzing the unscaled partition function $Z^w_{Y,\Lambda,a}$ of Eq. (\ref{Zuw}) and show how the important singular scaling factor $[g^2a^{4-d}]^{d(N)\Lambda_r/2}$ in the free energy is extracted. We note the same factor is extracted in both the upper and lower stability bound.

In the proof of Theorem \ref{thm2}, the following elementary bounds, considering the real function $(1-\cos u)$, are used:
\bequ\lb{ucos}
1\,-\,\cos u\,\leq\, \dfrac{u^2}2\quad , \quad u\in\mathbb R\,,
\eequ
and (see e.g. Ref. \cite{Simon3})
\bequ\lb{lcos}
1\,-\,\cos u\,\geq\, \dfrac{2u^2}{\pi^2}\qquad,\qquad |u|\leq \pi \,.
\eequ

Also, for $A=U$, $U$ unitary, $\|U\|_{H-S}\,=\,\left\{{\mathrm Tr}[U^\dagger U]\right\}^{1/2}\,=\,N^{1/2}$. The Euclidean induced norm of $U$ is one.
We also have the matrix equivalent norm inequalities
\bequ\lb{normineq}
N^{-1/2}\,\|A\|_{H-S}\,\leq\,\|A\|\,\leq\,\|A\|_{H-S}\,.
\eequ
The upper bound is well known. The lower bound follows by summing over $j$ the inequality $\|A\|^2\geq |\nu_j|^2$, where $\nu_j^2$ is an eigenvalue of the positive self-adjoint $N\times N$ matrix $A^\dagger A$.

Below, we describe how to obtain the upper and lower stability bounds on the scaled partition function $Z_{Y,\Lambda,a}$ appearing in Eq. (\ref{ZYn}).\vspace{2mm}

\noindent{\bf \underline{Upper Stability Bound}:}\vspace{3mm}

We describe how the upper stability bound is obtained. First, we recall that the gauge is not fixed, so that the enhanced temporal gauge is not used to obtain the upper bound. We note that, according to the definition of a class function given before, the action is {\em not} a class function of each bond variable in $\mathcal G$. To obtain the upper stability bound, we use the pointwise positivity of the Wilson action $(a^{d-4}/g^2)\mathcal A_p$ of each plaquette $p$.

The plaquettes perpendicular to the time direction $x^0$ are called {\em horizontal} plaquettes. By neglecting the horizontal plaquettes in the interior of the hypercubic lattice $\Lambda$ in $\mathcal S^w_{Y,\Lambda,a}=(a^{d-4}/g^2)\sum_p\,\mathcal A_p$ gives a lower bound on  $S^w_{Y,\Lambda,a}$ and an upper bound on $Z^w_{Y,\Lambda,a}$ of Eq. (\ref{Zuw}). Also, certain horizontal plaquettes having a bond in the boundary of $\Lambda$ are omitted.

After doing this, we integrate over successive layers of horizontal bonds, starting from the bonds associated with bond variables with the largest $x^0$ values and ending with the smallest values of $x^0$. In each step, the horizontal bond integration variable appears in only one plaquette. Using the left and right invariance of the Haar measure \cite{Simon2,Hall,Far}, and after integration of the horizontal bond variables, the remaining integral is independent of other bond variables of the plaquette. After completing the horizontal bond integration, the integrand is independent of the vertical bond variables. In this way, we obtain the upper bound $Z^w_{Y,\Lambda,a}\leq z^{\Lambda_r}$ It is remarkable and surprising that the bound factorizes into a product of single plaquette, single bond partition functions. Also, even though we have not fixed the gauge, the number of factors is $\Lambda_r$, the number of retained bonds in the enhanced temporal gauge.
We refer to \cite{YM} for more details, where an explicit example is treated for $d=3$.

It remains to obtain an upper bound on $z$
$$
z\,=\,\dis\int\,\exp\left(-\dfrac{a^{d-4}}{g^2}\,\|U-1\|^2_{H-S}\right)\,d\sigma(U)\,,
$$
where $d\sigma(U)$ is the Haar measure of the single-bond variable $U$.

We still need to control the single bond gauge integral for all values of the group element $g$. Here, we are in the pleasant situation where the single-bond action is a class function. This is where the Weyl integration formula \cite{Weyl,Metha,Simon2,Deift,Bump} enters! The group integration, originally involving a $N^2$-dimensional manifold integration over $\mathcal G$, reduces, for a class function, to an integral over the Haar measure of the $N$ angular eigenvalues $\la=(\la_1,\ldots,\la_N)$, $\la_j\in(-\pi,\pi]$, of a group element as given in Eq. (\ref{z}).

In the angular eigenvalue representation, an upper bound for $z$ is obtained by using a global upper bound for the Haar measure eigenvalue density $\rho(\la)$ given by
$$
\rho(\la)\,=\,\prod_{1\leq j<k\leq N}\,|e^{i\la_j}-e^{i\la_k}|^2\,=\,\prod_{1\leq j<k\leq N}\,\left\{2\,\left[1-\cos(\la_j-\la_k)\right]\right\}\,\leq\,\prod_{1\leq j<k\leq N}\,\left(\la_j-\la_k\right)^2\,\equiv\,\hat\rho(\la)\,,
$$
and a global lower bound for the single-bond action $(a^{d-4}/g^2)\,\mathcal A_s=(a^{d-4}/g^2)\,\|U-1\|^2_{H-S}$ in the exponent of Eq. (\ref{z}). The group global lower bound on $\mathcal A_s$ is $$\mathcal A_s\,=\,\|U-1\|_{H-S}^2\,=\,2\,\sum_{j=1,\ldots,N}(1-\cos\la_j)\,\geq\,\sum_{j=1,\ldots,N} (4\la_j^2)/\pi^2\,,$$ where $U$ is unitarily equivalent to ${\mathrm diag}(e^{i\la_1},\ldots,e^{i\la_N})$.

The upper bound on the density is a polynomial in the eigenvalues and each term is a monomial of the same degree, namely, $N(N-1)$. The Lebesgue measure has $N$ factors so, in making a scale transformation, we pick up an overall $N^2$ factor. An additional integration constraint is added for the ${\mathrm SU}(N)$ case, so that the product of the eigenvalues is one. After a change of variables, the scale factor for $z$ emerges and, hence, for $z^{\Lambda_r}$. Also, the resulting integral turns out to be proportional to the probability distribution of the Gaussian Unitary Ensemble (GUE) of random matrix theory. The integral is finite and can be evaluated explicitly. In this way, we obtain the upper stability bound.

To be explicit, recalling $\mathcal N_C(N)\,=\,(2\pi)^N\,N!$, we have
$$
z\,\leq\,\dfrac{1}{\mathcal N_C(N)}\,\dis\int_{(-\pi,\pi]^N}\, \exp\left[-4a^{d-4}/(g^2\pi^2)\,\sum_{1\leq j\leq N}\,\la_j^2\right]\,\hat\rho(\la)\,d^N\la\,.
$$

Changing variables to $y_j\,=\,\left[2a^{(d-4)/2}/(\pi g)\right]\,\la_j$, we get
$$
z\,\leq\,\dfrac{\left(g^2\,a^{4-d}\right)^{N^2/2}}{\mathcal N_C(N)}\,\left(\dfrac {\pi^2}{2}\right)^{N^2}\,\mathcal N_G(N)\,,
$$
where the integral is given below Eq. (\ref{I}).

Hence, as $Z^w_{Y,\Lambda,a}\,\leq\,z^{\Lambda_r}$, it follows that the scaled pure-gauge partition function of Eq. (\ref{ZYn}) obeys the bound
$$
Z_{Y,\Lambda,a}\,\leq\,\left[%\left(g^2\,a^{4-d}\right)^{N^2/2}\,
\dfrac{\mathcal N_G(N)}{\mathcal N_C(N)}\,\left(\dfrac {\pi^2}{4}\right)^{N^2}\,\right]^{\Lambda_r}\,.
$$
Hence, it suffices to take $c_{Y,u}\geq \ln \{[\pi/(2\sqrt{2})]^{N^2} I(\infty)\}$ in Theorem \ref{thm2}, where $I(\infty)$ is defined after Eq. (\ref{I}).

For $d=2$, the thermodynamic limit of the normalized free energy $f^n_{a,\Lambda}$ exists and, by dominated convergence, the continuum limit also exists and is $$f^n=-\ln\sqrt{2}-\dfrac 1{2N}\ln(2\pi)+\dfrac 1{N^2}\sum_{1\leq j\leq(N-1)}\ln (j!)\,.$$\vspace{2mm}

\noindent{\bf \underline{Lower Stability Bound}:} \vspace{3mm}

To derive the lower bound on the unscaled partition $Z^w_{Y,\Lambda,a}$ of Eq. (\ref{Zuw}), we fix the enhanced temporal gauge. The number of retained bonds is $\Lambda_r\simeq(d-1)L^d$, for $L\gg 1$. Depending on the location and orientation of a plaquette, there can be one, two or four retained bond variables which are to be integrated. We have proved (see Lemma \ref{lquad} below), for a single plaquette action, a global quadratic bound in the plaquette's gluon field bond variables. Using this bound, the partition function $Z^w_{Y,\Lambda,a}$ factorizes over the retained bond variables. Since each bond is present in at most $(d-1)$ plaquettes, we use an overcounting argument to replace the sum over plaquette actions by a sum over quadratic gluon field bond actions. With this, $Z^w_{Y,\Lambda,a}$ is bounded below by $z_1^{\Lambda_r}$, where $z_1$ is the partition function with a single bond variable quadratic action. This single bond variable quadratic action is a class function on the gauge group $\mathcal G$. Hence, the Weyl integration formula can be applied and the group integral reduces to an integral over the angular eigenvalues. We let $\tilde z$ denote the integral of $z_1$ with the integration domain restricted so that the bond variable is near the identity. Thus, $z_1\geq \tilde z$ and $Z^w_{Y,\Lambda,a}\geq \tilde z^{\Lambda_r}$. Next, we use a lower bound on the Haar measure density $\rho(\la)$. Last, changing variables in the retained bond variables, we obtain the lower stability bound of Theorem \ref{thm2} with the same factor $\Lambda_r$ that occurs in the upper stability bound, namely $(g^2a^{4-d})^{N^2\Lambda_r/2}$.

The proof that follows is a very simplified version of the proof presented in Ref. \cite{YM}.

We first give some facts about a unitary matrix and, for the gauge group $\mathcal G=\mathrm U(N)$, also for the exponential map from the Lie algebra to the group.
\begin{lemma}\lb{diag}
Take $\mathcal G=U(N)$. Given a matrix $U\in\mathcal G$, with eigenvalues $e^{i\la_1}$, ..., $e^{i\la_N}$, and angular eigenvalues $\la_1$, ..., $\la_N\in(-\pi,\pi]$, there exists a unitary matrix $V\in\mathcal G$ that diagonalizes $U$, i.e. $V^{-1}UV\,=\,{\mathrm diag}(e^{i\la_1},\ldots , e^{i\la_N})$. Set $X\,=\,V\,{\mathrm diag}(\la_1,\ldots , \la_N)\,V^{-1}$ and $x_\alpha={\mathrm Tr} (X\theta_\alpha)$, so that $X\,=\,\sum_{\alpha=1,...,d(N)}\,x^\alpha\theta_{\alpha}$. Then, for $x=(x^1,\ldots,x^{d(N)})$ and $\la=(\la_1,\ldots,\la_{d(N)})$, we have $U\,=\,e^{iX}$ and $\|X\|^2_{H-S}\,=\,|x|^2\,=\,|\la|^2\,\leq\,  N\pi^2$. The exponential map, with $|x|^2\leq N\pi^2$, is onto the group $\mathcal G$.
\end{lemma}

\noindent{\bf Proof of Lemma \ref{diag}: } The first part is just the spectral theorem for a unitary matrix. Calculating $e^{iX}$, with $X=V\,{\mathrm diag}(\la_1,\ldots , \la_N)\,V^{-1}$ shows that $U=e^{iX}$ so that the exponential map is onto \cite{Simon2,Hall,Far}. Similarly, calculating $\|X\|^2_{H-S}$ shows that  $\|X\|^2_{H-S}=|\la|^2$. Thus, the exponential map is onto for $|x|^2\leq N\pi^2$. On the other hand, calculating $\|X\|^2_{H-S}$ from the representation $X=\sum_{\alpha=1,\ldots,d(N)}\,x^\alpha\theta_{\alpha}$ gives  $\|X\|^2_{H-S}=|x|^2$, so that $|x|^2=|\la|^2$. \qed 
\begin{rema}
We note that with the domain restriction $\|U-1\|<1$ we have a well defined self-adjoint $X=-i\ln [1+(U-1)]=-i\sum_{j\geq 1}\,(-1)^{j+1}\,(U-1)^j/j$. Also, without the domain restriction, except for some special cases such as $\mathcal G={\mathrm SU}(2)$, we do not know the minimal domain of $X_b$ such that  the $e^{iX_b}$'s cover the whole gauge group.
\end{rema}

We now give a new global quadratic upper bound on a generic plaquette action $(a^{d-4}/g^2)\mathcal A_p$, for $\mathcal A_p\,=\,\|U_p-1\|^2_{H-S}$ and on the Wilson plaquette action $S^w_{Y,\Lambda,a}\,=\,(a^{d-4}/g^2)\,\sum_p\,\mathcal A_p$ [see Eq. (\ref{YMaction})]. Here, recall that $U_p=U_1U_2U^\dag_3U^\dag_4$ and $U_\ell=e^{iX_\ell}$, $\ell=1,2,3,4$, where $X_\ell$ is defined as in Lemma \ref{diag}, so that $\|X_\ell\|^2_{H-S}\,=\,|x^\ell|^2$. With this, we have:
\begin{lemma}\lb{lquad}
For the four retained bond plaquette, we have the global quadratic upper bound on $\mathcal A_p$:
\bequ\lb{lower1}
\mathcal A_p\,=\,\|U_p-1\|^2_{H-S}\,\leq\, C^2 \,\sum_{1\leq j\leq 4}\, |x^j|^2\,=\, C^2 \,\sum_{1\leq j\leq 4}\, |\la_j|^2\quad\:\:,\:\:\quad C=2\sqrt N\,.
\eequ
When there are only one, two or three retained bond variables in a plaquette, the sum over $j$ has, respectively, only one, two and three terms and the numerical factor $4$ in $C^2=4N$ is replaced by 1, 2 and 3, respectively. For the total action $S^w_{Y,\Lambda,a}\,=\,(a^{d-4}/g^2)\,\sum_p\,\mathcal A_p$, we have the global quadratic upper bound
\bequ\lb{lower2}S^w_{Y,\Lambda,a}\,\leq\,2(d-1)\,C^2\,(a^{d-4}/g^2)\,\sum_b|x^b|^2\,=\,2(d-1)\,C^2\,(a^{d-4}/g^2)\,\sum_b|\la_b|^2\,,
\eequ
where the sum runs over all $\La_r$ retained bonds.
\end{lemma}
\begin{rema}\lb{remake}
The total upper bound of Eq. (\ref{lower2}) results from Eq. (\ref{lower1}) observing that a lattice bond is present in at most  $2(d-1)$ plaquettes.
	\end{rema}
\begin{rema}
	The coefficient $C$ does not depend on the size of the gluon fields, such as, the upper bound on $\mathcal A_p$ holds also for large fields, in contrast to the Lemma proved in \cite{YM}, where $C^2$ grows quadratically in the fields. Of course, $\|U_p-1\|^2_{H-S}\,=\,{\mathrm Tr} [2-U_p-U^\dagger_p ]\,\leq\,4N$, is a global constant upper bound, but we use the quadratic behavior coupled with a scale change of variables to extract the dominant $a$ and $g^2$ behavior from the partition function $Z^w_{Y,\Lambda,a}$.
\end{rema}
\begin{rema}
	We emphasize that the new global upper bound is quadratic in the fields even though the Riemann sum approximation to the classical smooth field continuum action in Eq. (\ref{TrF2}), obtained by a formal small $a$ expansion, has local cubic and (positive) quartic terms.
\end{rema}

The proof of Eq. (\ref{lower1}), and then of Lemma \ref{lquad}, is deferred to the end of the section.

Using in the exponent action the lower bound of Lemma \ref{lquad} and the Weyl integration formula for each of the $\Lambda_r$ retained bonds, we obtain the lower bound
$$
Z^w_{Y,\Lambda,a}\,\geq\, z_1^{\lambda_r}\,,
$$
where
$$
z_1\,=\,\dfrac1{\mathcal N_c(N)}\,\int_{(-\pi,\pi]}\;\exp\left[-\,\dfrac{a^{d-4}}{g^2}\,2(d-1)C^2|\la|^2\right]\,\rho(\la)\,d^N\la\,.
$$

We now use a lower bound on the measure density $\rho(\la)$. It is here that we impose small field restrictions. Since $(1-\cos u)\geq 2(u^2/\pi^2)$, $|u|\leq\pi$, we have $$|e^{i\la_j}-e^{i\la_k}|\,=\,2[1-\cos(\la_j-\la_k)]\,\geq 
\, \dfrac4{\pi^2}\,\left(\la_j-\la_k\right)^2\qquad,\qquad |\la_\ell|\leq\dfrac\pi 2\quad,\quad\ell=j,k\,,
$$
and
$$
\rho(\la)\,=\,\prod_{1\leq i<j\leq N}\,|e^{i\la_j}-e^{i\la_k}|^2\,\geq\,\left(\dfrac4{\pi^2}\right)^{N(N-1)/2}\,\hat\rho(\la)\qquad,\qquad |\la_\ell|\leq\dfrac\pi 2\,.
$$
Thus, with this, we have $z_1\,\geq\,\check z$ where
\bequ\lb{checkz}
\check z\,=\,\dfrac 1{\mathcal N_C(N)}\,\left(\dfrac4{\pi^2}\right)^{N(N-1)/2}\,\int_{|\lambda_k|\leq\pi/2}\,e^{-2c(d-1)C^2\sum_{k=1}^N \lambda^2_k}\hat\rho(\lambda)d^N\lambda\,,
\eequ
where $c\equiv c(a,g^2,d)=a^{d-4}/g^2$.

So far, we have $Z^w_{Y,\Lambda,a}\,\geq\,\check z^{\lambda_r}$. Making the change of variables $y_\ell\,=\,[2(d-1)C^2a^{d-4}/g^2]^{1/2}\la_\ell$, we obtain
$$
\check z\,=\, \dfrac1{\mathcal N_C}\,\left(\dfrac{a^{d-4}}{g^2}\right)^{-N^2/2}\,\left(\dfrac4{\pi^2}\right)^{N(N-1)/2}\left[ 2(d-1)C^2\right]^{-N^2/2}\;I([2(d-1)C^2a^{d-4}/g^2]^{1/2}\pi/2)\,, 
$$
with $I(u)$ being the integral defined by Eq. (\ref{I}).

By the monotonicity of $I(u)$ it assumes its minimum value $I_\ell>0$ for $a=1$ and at a finite $g^2=g^2_0>0$. Thus,
$$
\check z\,\geq\,\dfrac1{\mathcal N_C}\,\left(\dfrac{a^{d-4}}{g^2}\right)^{-N^2/2}\,\left(\dfrac4{\pi^2}\right)^{N(N-1)/2}\left[ 2(d-1)C^2\right]^{-N^2/2}\,I_\ell\,.
$$

In this way, the lower stability bound of Theorem \ref{thm2}, for the normalized Yang-Mills partition function $Z_{Y,\Lambda,a}\,=\,s_Y^{d(N)\Lambda_r}\,Z^w_{Y,\Lambda,a}$
$$
Z_{Y,\Lambda,a}\,\geq\,e^{c_\ell\Lambda_r}\,,
$$
holds for $c_\ell\,\leq\,\ln\left\{\mathcal N_c^{-1}\,\left(4/\pi^2\right)^{N(N-1)/2}\,\left[2(d-1)C^2\right]^{-N^2/2}\;I_\ell\right\}$.

Now, we go back to Lemma \ref{lquad} and prove the bound of Eq. (\ref{lower2}). According to Remark \ref{remake}, this is the missing point to accomplish the proof of Lemma \ref{lquad}.
Our proof here improves the result given in Lemma 1 of \cite{YM}. Below, we also consider the case of three retained bonds, which is absent in the enhanced temporal gauge that we use. For simplicity, here we take the case where we have four retained bonds in a plaquette. We define, for $1\leq j\leq 4$, $\mathcal L_j=i\sum_{\alpha=1,\ldots,d(N)} x^j_\alpha\theta_\alpha$, so that  $U_j=e^{\mathcal L_j}$ and $U_p=U_1U_2U_3^\dagger U^\dagger_4$.

First, we remark that $\|\mathcal L_j\|\,\leq\,\|\mathcal L_j\|_{H-S}\,=\,|x^j|$. Next, let $U_p(\delta)\,=\,U_1(\delta)U_2(\delta)U^\dagger_3(\delta)U^\dagger_4(\delta)$, $U_j(\delta)\,=\,e^{\delta\mathcal L_j}$, for $\delta\in[0,1]$. By the fundamental theorem of calculus, suppressing all $\delta$ dependence,
$$
U_p\,-\,1\,=\,\dis\int_0^1\,d\delta\,\left[\mathcal L_1U_1U_2U_3^\dagger U_4^\dagger\,+\,U_1\mathcal L_2U_2U_3^\dagger U_4^\dagger\,-\,U_1U_2\mathcal L_3U_3^\dagger U_4^\dagger\,-\,U_1U_2U_3^\dagger\mathcal L_4 U_4^\dagger\right]\,,
$$
and, by the triangle and Cauchy-Schwarz inequalities, it follows that
$$
\|U_p\,-\,1\|\,\leq\,\sum_{j=1}^4\,\|\mathcal L_j\|\,\leq\,\sum_{j=1}^4\,\|\mathcal L_j\|_{H-S}\,=\,\sum_{j=1}^4\,|x^j|\,\leq\,2\,\left[\sum_{j=1}^4\,|x^j|^2\right]^{1/2}\,.
$$
But, $\|U_p\,-\,1\|\,\geq\,N^{-1/2}\,\|U_p-1\|_{H-S}$. Hence
$$
\mathcal A_p\,=\,\|U_p-1\|^2_{H-S}\,\leq\,4N\,\sum_{j=1}^4\,|x^j|^2\,.
$$

From the above proof, we easily see that the numerical factor $4$ in $C^2$ is replaced by $1$, $2$ and $3$, respectively, when there are only one, two or three retained bond variables in a plaquette, since the sum over $j$ has one, two or three terms, respectively. The proof of Lemma \ref{lquad} is finished.\qed
%=============================================================
%=============================================================
%=============================================================
\subsection{Proof of Theorem \ref{thm4}}
%=============================================================
%=============================================================
%=============================================================
To prove Theorem \ref{thm4}, first, in the integral of Eq. (\ref{w4}), we make the change of variables $\la\,=\,\sqrt\beta y$ to get
$$\barr{lll}
w&=&\dfrac{\beta^{N/2}}{\mathcal N_c(N)}\,\dis\int_{(-\pi/\sqrt{\beta},\pi\sqrt\beta]^N}\,e^{-L(\sqrt{\beta}\la)/\beta}\,\prod_{1\leq j<k\leq N}|e^{i\sqrt{\beta} y_j}-e^{i\sqrt{\beta} y_k}|^2\,d^Ny\\
&\leq&\dfrac{\beta^{N^2/2}}{\mathcal N_C(N)}\,\dis
\int_{(-\pi/\sqrt{\beta},\pi/\sqrt{\beta}]^N}\,e^{-cy^2}\,\prod_{1\leq j<k\leq N}(y_j-y_k)^2\,d^Ny\,.\earr
$$
Here, we used $|e^{iu}-e^{iv}|^2\,=\,2[1-\cos(u-v)]\,\leq\,(u-v)^2$.

The positive integrand is integrable over $\mathbb R^N$, so that, by the Lebesgue dominated convergence theorem, the integral converges to the integral over $\mathbb R^N$. The $\mathbb R^N$ integral value is, for $c=1$, $\mathcal N_G(N)\,=\,(2\pi)^{N/2}\,2^{-N^2/2} \prod_{1\leq j\leq N}\,j!$ (see \cite{Metha,HC,IZ,ZJZ,Z}).
The proof of Theorem \ref{thm4} is then finished.\qed
%%%%%%%%%%%%%%%%%%%%%%%%%%%%%%%%%%%%%%%%%%%%%%%%%%%%%%%%%%%%%%%%%%%%%%%%%%
%%%%%%%%%%%%%%%%%%%%%%%%%%%%%%%%%%%%%%%%%%%%%%%%%%%%%%%%%%%%%%%%%%%%%%%%%%
\section{Concluding Remarks}\lb{sec4}
%%%%%%%%%%%%%%%%%%%%%%%%%%%%%%%%%%%%%%%%%%%%%%%%%%%%%%%%%%%%%%%%%%%%%%%%%%
%%%%%%%%%%%%%%%%%%%%%%%%%%%%%%%%%%%%%%%%%%%%%%%%%%%%%%%%%%%%%%%%%%%%%%%%%%
One of the main objectives in the analysis of a relativistic quantum field model is to prove its existence and determine its particle content and scattering.  Here, we consider complex and real multiflavored scalar, spin zero field Bose-gauge lattice models in an imaginary time functional integral formulation. The models are defined on a finite hypercubic lattice $\Lambda\subset a\mathbb Z^d$, with $L\in\mathbb N$ sites on a side,  lattice spacing $a\in(0,1]$ and for spacetime dimension $d = 2,3,4$ dimensions. The number of sites in the lattice $\Lambda$ is denoted by $\Lambda_s=L^d$. The lattice provides an ultraviolet regularization.

The model action has two terms: a pure-gauge, Yang-Mills term and a term with the Bose fields minimally coupled to the gauge fields. The Wilson plaquette action $(a^{d-4}/g^2)\sum_p\,\mathcal A_p$, where $p$ is a lattice minimal square or plaquette, is used for the pure-gauge action, for both abelian and non-abelian compact and connected gauge Lie groups $\mathcal G\,=\,\mathrm U(N)\,,\,\mathrm {SU}(N)$. The gauge variables $g_b$ are associated with the lattice bonds $b$, connecting nearest neighbor sites, and take matrix values in an irreducible unitary representation of $\mathcal G$. Each term in the action is gauge invariant under local gauge transformations with the gauge group $\mathcal G$. Boundary conditions play an important role and we adopt free boundary conditions. The naive $a\searrow 0$ or continuum limit of this action gives the smooth field classical continuum Yang-Mills action plus the minimal coupling between the gauge and Bose fields \cite{Gat}.

Here, we consider the problem of obtaining thermodynamic stable stability bounds for the original (called {\em unscaled}) model partition function $Z^u_{\Lambda,a}$, with the {\em same} singular factor for the free energy in the upper and lower bounds, as $a\searrow 0$. Once this singular factor is extracted from the unscaled partition function, using a multiplicative scaling factor and defining a new, 
scaled partition function $Z_{\Lambda,a}$, for which we show the finite-lattice model satisfies thermodynamic and also ultraviolet stable stability bounds. The bounds on $Z_{\Lambda,a}$ are new. They arise, in part, from a new upper bound on the Wilson plaquette action. The upper bound is local, global and quadratic in the gluon bond variables. The bound is surprising since the naive small $a$ approximation to the action has cubic and positive quartic terms in case of a non-abelian gauge group. 

Regarding the structure of our stability bounds, for pure Yang-Mills models the bound factorizes; the number of factors is roughly the number of lattice bonds in the temporal direction, i.e. $[(d-1)L^d]$. Each factor is a partition function with a single plaquette action and a single-bond gauge variable. Furthermore, each factor is bounded by an integral over the Gaussian unitary random matrix ensemble probability distribution (GUE).

For the Bose-gauge partition function, the bound also factorizes. The number of factors is approximately the total number of lattice bonds, i.e. $dL^d$. Each factor is a partition function with a single-bond action and two Bose field variables. Similar factorization occurs for the generating function for plaquette Bose fields with a source term, like the case analyzed in \cite{Schor}. The source field is uniform over the lattice. Although in this paper we assumed a small field restriction \ch (not needed for the gauge group $\mathrm U(1)$), no small or large field restriction are needed to obtain the factorized bounds.

Coming back to the scaled partition function $Z_{\Lambda,a}$, we have the thermodynamic and also ultraviolet stable stability bounds of the type
$$
e^{c_{\ell}\Lambda_s} \leq Z_{\Lambda,a} \leq e^{c_{u}\Lambda_s}\,,
$$
with real constants $c_\ell$ and $c_u$ independent of the size $L$ of $\Lambda$ and $a$. Hence, by the Bolzano-Weierstrass theorem, the above stability bounds imply the existence of the thermodynamic limit ($\Lambda\nearrow a\mathbb Z^d$) and, subsequently, the continuum limit ($a\searrow 0$) of the scaled or normalized free energy $f^n_{\Lambda,a}\equiv\ln Z_{\Lambda,a}/\Lambda_s$, at least in the subsequential sense. The existence of the scaled free energy in the thermodynamic and continuum limits, arising from the knowledge of the exact singular behavior of the unscaled field free energy, is the only problem addressed here. We do {\em not} investigate any other physical properties of the limiting models e.g. the particle spectrum of the model, its correlations, scattering, etc.

Our stability bounds emerge from the use of scaled fields. The scaling is {\em noncanonical\;}! For the Bose fields, $a$-dependent scaled local Bose fields are used at the outset, and the scaling is like a wavefunction renormalization. For the gauge fields, the introduction of the scaling is more complicated. First, we use the gauge group exponential map to pass from the bond variables in $\mathcal G$ to the gauge, gluon fields or potentials associated with the parameters of its Lie algebra. Then, the scaling is performed on the gluon fields.

By a gauge transformation, sometimes it is convenient to fix the {\em enhanced} temporal gauge and eliminate the excess of gauge degrees of freedom from the problem. In the enhanced temporal gauge, the bond variables in the temporal direction are set
to the identity element of $\mathcal G$, and the gauge integrals dropped. There are also specified gauge bond variables on the boundary of the lattice $\Lambda$ that are similarly set to the identity. The gauged away bond variables $g\in\mathcal G$ are related to bonds of a maximal tree in $\Lambda$. The remaining bonds and bond variables are called {\em retained} bonds and retained bond variables, respectively. We denote the number of retained bonds by $\Lambda_r$ and we have that $\Lambda_r\simeq (d-1)\Lambda_s$, if $L\gg 1$. The enhanced temporal gauge fixing does not alter the value of the partition function $Z^u_{\Lambda,a}$ as there are no loops with identity elements on the lattice bonds. Only a maximal tree in the lattice $\Lambda$ is involved.

Using a simple generic bound on the integral of the product of two functions,  our stability bounds for the scaled partition function $Z_{\Lambda,a}$ are obtained by showing thermodynamic and ultraviolet stable bounds separately for the scaled partition function $Z_{Y,\Lambda,a}$ of the pure-gauge Yang-Mills model and the scaled partition function $Z_{B,\Lambda,a}(\tilde g)$ of the model with Bose fields minimally coupled to the gauge fields. For the latter, we postpone the gauge field integration over the retained gauge variables $\tilde g$. It turns out that our stability bounds on $Z_{B,\Lambda,a}(\tilde g)$ are uniform in $\tilde g$ so that postponing the retained gauge integrals does not introduce any problem! This is similar to  what happens when the diamagnetic inequality is applied \cite{BFS}.

Our choice of scaling for the gauge fields allows us to extract easily and directly the precise singular factor of the free energy. There is no need to employ e.g. complex multiscale methods like the renormalization group method, as done in \cite{Bal,Bal2}. A multiplicative scaling factor is used to define the scaled pure-gauge partition function $Z_{Y,\Lambda,a}$ so that the associated normalized free energy exists and is bounded in the thermodynamic and continuum limits.

We point out that the pure-gauge and the Bose-gauge partition functions are treated independently and are, separately, both thermodynamic and ultraviolet stable.

To be more precise and see how all this works in more detail, we mention that the Bose integrated partition function with an arbitrary retained gauge field configuration is shown to obey stability bounds. The proof of these bounds is new, holds also for more general lattices, and bypasses diamagnetic inequality considerations. The proof uses H\"older's inequality in the partition function to obtain bounds which decouple the $d$ coordinate directions. The bound on the scaled partition function is reduced to a product of bounds on one-dimensional chain partition functions. The number of factors or number of chains is $L^{d-1}$. The bound on the partition function for a chain is further factorized and reduces to a product (over the bonds of the chain) of bounds on a single bond `transfer matrix'. Thus, bounds on the arbitrarily large lattice partition function are factorized and reduced to a bound on an almost local object, namely, the single bond `transfer matrix'.
Also, since the scalar Bose field integral is Gaussian, the integral gives the negative square root of the determinant of a quadratic form matrix. Thus, there is an interplay between H\"older's inequality, determinantal inequalities and the spectrum of each summand in a decomposition of the quadratic form matrix, which is an avenue open for further exploration. Our stability bounds on the scaled Bose-gauge partition function also hold when the Bose field bare mass is zero. The use of free boundary conditions is crucial to ensure stability. If e.g periodic boundary conditions are adopted, we lose stability. This is a consequence of the presence of zero modes contributing to the Bose quadratic form in the action.

For the pure-gauge partition function for the group $\mathcal G=\mathrm U(N)$ (respectively, $\mathrm{SU}(N)$), there are $N^2$ (respectively, $N^2-1$) gluon fields. Each component of the gauge field is taken to locally scale like the massless scalar Bose field, namely, with the factor $a^{(d-2)/2}$. Formally, the gluon scaling factor gives rise to the scaling factor for the unscaled Wilson partition function $Z^w_{Y,\Lambda,a}$.
Using this scaling factor, the scaled pure-gauge partition function $Z_{Y,\Lambda,a}$ is defined and is proved to satisfy thermodynamic and ultraviolet stable stability bounds.

Similar to the bounds for the Bose-gauge partition function $Z_{B,\Lambda,a}(\tilde g)$, the bounds on the scaled pure-gauge partition function $Z_{Y,\Lambda,a}$ follow from bounds on $Z^w_{Y,\Lambda,a}$. The bounds on $Z^w_{Y,\Lambda,a}$ factorize and are reduced to bounds on almost local objects. For the lower bound, after fixing the enhanced temporal gauge on $Z^w_{Y,\Lambda,a}$, we have a new pointwise global {\em quadratic} upper bound in the gluon fields for the single Wilson plaquette action with one, two, three or four bond variables.  Using this and restricting the group bond variables to be near the identity, we obtain $Z^w_{Y,\Lambda,a}\geq\tilde z^{\Lambda_r}$, where $\tilde z$ is a single plaquette, single bond partition function. For the upper bound on $Z^w_{Y,\Lambda,a}$, we do not use any gauge fixing. Also, we do not have a pointwise quadratic lower bound on the single plaquette action. But, using the pointwise positivity of each plaquette action and eliminating the interior horizontal plaquettes plus some additional horizontal plaquettes on the boundary, we have a lower bound on the total action and an upper bound on $Z^w_{Y,\Lambda,a}$. Here, again, the bound factorizs and the upper bound is given by $z^{\Lambda_r}$, where $z$ is a single plaquette partition function with a single gauge bond variable. In this case, there is a global quadratic lower bound for the plaquette action in each $z$ and a global upper bound on the single variable Haar measure.

We still need to bound $z$ from above and $\tilde z$ from below. We observe that the original Wilson partition function $Z^w_{Y,\Lambda,a}$ is not a class function of each retained bond variable in $\mathcal G$. However, $z$ and $\tilde z$ are class functions of a single bond variable.  The Weyl integration formula applies and replaces the gauge group integration by the integration over the angular eigenvalues of a group element. For $\mathcal G=\mathrm U(N)$, the unitary group element has eigenvalues $e^{i\la_1}$, ..., $e^{i\la_N}$, $\la_j\in(-\pi,\pi]$, and $\la_1$, ..., $\la_N$ are the angular eigenvalues. The $N^2$-dimensional integration over the gluon fields is reduced to an $N$-dimensional integration over the associated angular eigenvalues, with a density times a Lebesgue measure. The density can be easily bounded from above and below.

For the upper bound on $Z_{Y,\Lambda,a}$, we use a global lower bound on the action of $z$. For the lower bound on $Z_{Y,\Lambda,a}$, on the action of $\tilde z$, we use a new global upper bound. Both bounds are local and quadratic! And, with these bounds, the same singular scaling factor is extracted from both the upper and lower bounds on $Z^w_{Y,\Lambda,a}$. The fact that the same singularity is extracted from both the lower and upper bounds is important as it allows us to define the scaled pure-gauge partition using a multiplicative scaling factor. It is this function that obeys the thermodynamic and ultraviolet stable stability bounds. The stability bounds hold even though it is not known if the mass gap persists in the limit $a\searrow 0$.

For the lower bound on the action of $\tilde z$, we obtain a new upper bound on the Wilson four-bond single plaquette action $(a^{d-4}/g^2)\,\mathcal A_p$ in terms of the gluon fields. The quadratic bound on $\mathcal A_p$  is global, meaning that it holds for all values of the gluon fields. The lower bound on $z$ emerges after using this upper bound on $\mathcal A_p$ and reducing the integration domain for the angular eigenvalues to provide a lower bound for the eigenvalue density of the angular eigenvalue Haar measure.

It is to be emphasized that the upper bound on the single plaquette action $(a^{d-4}/g^2)\,\mathcal A_p$ is quadratic in the gluon fields while, in the non-abelian case, the Riemann sum, in the naive small $a$ approximation to the classical continuum spacetime smooth field Yang-Mills action has cubic and quartic terms. The quartic term is local and positive. In this approximation, we observe that the quartic term has {\em at most} two identical (same components) fields at a lattice site. This gives more regularity in the continuum limit, as compared to e.g. the $P(\phi)$ scalar models where we can have the product of four or more identical Bose fields at a same point \cite{Riv}.

In the special case of the gauge group ${\mathrm SU}(2)$, there is a widely used, nice geometric description of the group as there is a $1-1$ correspondence between the group and the unit sphere $S^3$ in $\mathbb R^4$ \cite{Simon2,Hall}. In terms of the gluon field parametrization of the group, an explicit formula for the Haar measure is available using the restriction of spherical coordinates in $\mathbb R^4$ to $S^3$. A direct proof of thermodynamic and ultraviolet stable stability bounds in terms of the gluon fields is given here. On the other hand, the Weyl integration formula with the parametrization of the group in terms of angular eigenvalues also applies. We show that, for ${\mathrm SU}(2)$, the single plaquette partition function with a single bond variable expressed in terms of gluon fields is the same as the Weyl angular eigenvalue formula.

Our methods give the exact result for $d = 2$. In this case, besides the  stability bounds, we also obtain explicitly the continuum limit ($a\searrow 0$) of a normalized free energy, and a connection is made between the Circular Unitary Ensemble (CUE) and the Gaussian Unitary Ensemble (GUE) of random matrix theory (see \cite{Metha,Deift,AGZ}). This connection between a Haar measure representation of group theory and random matrix theory appears in a natural way. The representation for the continuum limit is the GUE probability distribution for eigenvalues of self-adjoint matrices in the case of the gauge group $\mathcal G={\mathrm U}(N)$. Similar results and representations hold for the other classical groups.

The same type of analysis can also be made when fermion fields, like quarks and antiquarks in QCD, are present. Indeed, neglecting the pure-gauge part, the fermion-gauge part was treated in Refs. \cite{MP,MP2}, and an upper stability bound was also obtained, independent of the lattice spacing $a$. However, Jensen's inequality, used here to derive a lower stability bound for the Bose-gauge part, does not work with fermions. We need to find an alternative method. Once we can overcome this difficulty, combining the present results with the ones of Refs. \cite{MP,MP2}, we can start considering the problem of showing stability bounds and the existence of a normalized free energy for QCD, in the thermodynamic and continuum limits.

Despite being fundamental, all these questions about stability do not give information on the energy-momentum spectrum, local clustering properties and the model particle content. For lattice QCD, with fixed unit lattice spacing and working in the strong coupling regime, where the gauge coupling is much smaller than a small hopping parameter, we also have results validating the Gell'Mann-Ne'eman `eightfold way', the exponential decay of the meson-like exchange Yukawa interaction and the existence of some hadron-hadron bound states (see e.g Ref. \cite{WPT-Baryons,WPT-Mesons} and references therein). It would be nice to have some of the general properties and laws of nuclear physics rigorously derived from first principles, i.e. from fundamental quark and gluon fields and QCD dynamics based on the gauge, color group ${\mathrm SU}(3)$.
%%%%%%%%%%%%%%%%%%%%%%%%%%%%%%%%%%%%%%%%%%%%%%%%%%%%%%%%%%%%%%%%%%%%%%%%%%%%
%%%%%%%%%%%%%%%%%%%%%%%%%%%%%%%%%%%%%%%%%%%%%%%%%%%%%%%%%%%%%%%%%%%%%%%%%%%%
%%%%%%%%%%%%%%%%%%%%%%%%%%%%%%%%%%%%%%%%%%%%%%%%%%%%%%%%%%%%%%%%%%%%%%%%%%
\appendix{\begin{center}{\bf APPENDIX A: Complex Bose Field Case}\end{center}}
%%%%%%%%%%%%%%%%%%%%%%%%%%%%%%%%%%%%%%%%%%%%%%%%%%%%%%%%%%%%%%%%%%%%%%%%%%%%%%
\lb{appA}
\setcounter{equation}{0}
\setcounter{lemma}{0}
\setcounter{thm}{0}
\renewcommand{\theequation}{A\arabic{equation}}
\renewcommand{\thethm}{A\arabic{thm}}
\renewcommand{\thelemma}{A\arabic{lemma}$\:$}
\renewcommand{\therema}{{\em{\bf A\arabic{rema}$\:$}}}\vspace{.5cm}
%%%%%%%%%%%%%%%%%%%%%%%%%%%%%%%%%%%%%%%%%%%%%%%%%%%%%%%%%%%%%%%%%%%%%%%%%%%%%
%*************************************************************************
%%%%%%%%%%%%%%%%%%%%%%%%%%%%%%%%
Here, we generalize our analysis in showing Theorem \ref{thm1} for real Bose fields to complex fields. In the hopping term in the gauge-matter action, with $\phi\equiv \phi(x_\mu)$ and $\phi^+\equiv \phi(x^+_\mu)$,we have
$$
\bar \phi g \phi^+ \,+\, \bar \phi^+ g^{-1} \phi\,.
$$
Letting $\phi\,=\,\phi_R\,+\,i\phi_I\,=\,R\,+\,iI$, $d\tilde\phi(x)\,=\,d\tilde R(x)\,d\tilde I(x)$, we have
$$
\bar \phi g \phi^+ \,=\,\left(\barr{cc}  R&I \earr\right)
\;\left(\barr{cc}g&ig\\-ig&g \earr\right)\;\left(\barr{c}R^+\\I^+  \earr\right)\,\equiv\,\left(   \vec{\phi},M\vec\phi^{\,+}\right)\,,$$
where $M$ is given by the above $2\times 2$ matrix, and we use the Euclidean inner product in $\mathbb R^{2N}$, with $\vec{\phi}\,=\,(R,I)$. Notice that, since $g$ is in a unitary representation of the gauge group $\mathcal G$, the second term in the action exponential is the complex conjugate of the first.

After applying the H\"older's inequality, like our treatment for the real Bose field case (check Eq. (\ref{kerT})), the one-bond transfer matrix is (c.c. denotes the complex conjugate)
$$
e^{-\frac14 |\vec\phi|^2}\;\exp\left\{\frac{d\kappa^2}{2}\, \left[\left(\vec\phi, M\vec{\phi}^+\right)\, +\,{\mathrm c.c.}\right]\right\}\;e^{-\frac14 |\vec\phi^+|^2}\,.
$$

We calculate the Holmgren bound for $d\kappa^2\,\rightarrow\,\frac12$, when $a\searrow 0$. Letting $L=M+\bar{M}$, we get
$$
\sup_{\phi}\;\left\{ e^{-\frac14 |\vec\phi|^2}\;\dis\int\,\exp\left[\frac{1}{4}\, \left(\vec\phi, L\vec{\phi}_2\right)\right]\,dR_2\,dI_2\,\;e^{-\frac14 |\vec\phi_2|^2}\right\}\,,
$$
from which, upon performing the Gaussian integration, we obtain
$$
2^N\,\sup_{\phi}\;\left\{ e^{-\frac14 |\vec\phi|^2}\;e^{\frac1{16}\, \left(L\vec\phi, L\vec{\phi}\right)}\right\}\,.
$$
For $^t$ denoting the transpose, we have $M^tM\,=\,0$, $\bar{M}^tM\,=\,2\left(  \barr{rr}  \bar{g}^tg\,&i\bar{g}^tg\\-i\bar{g}^tg\,&\bar{g}^tg  \earr\right)$ so that, using the fact $g$ is unitary, $L^tL\,=\,[M^tM+\bar{M}^tM]\,+\,{\mathrm c.c.}\,=\,4$. The integral in the Holmgren bound is then bounded by $2^N$. The Holmgren bound is also finite for a general matrix $g$ if $|g|\leq 1$ and $\bar{g}^tg$ is real.
% &&&&&&&&&&&&&&&&&&&&&&&&&&&&&&&&&&&&&&&&&&&&&&&&&&&&&&&&&&&&&&&&&&&&&&&
% &&&&&&&&&&&&&&&&&&&&&&&&&&&&&&&&&&&&&&&&&&&&&&&&&&&&&&&&&&&&&&&&&&&&&&&
%%%%%%%%%%%%%%%%%%%%%%%%%%%%%%%%%%%%%%%%%%%%%%%%%%%%%%%%%%%%%%%%%%%%%%%%%%%%
%%%%%%%%%%%%%%%%%%%%%%%%%%%%%%%%%%%%%%%%%%%%%%%%%%%%%%%%%%%%%%%%%%%%%%%%%%%%
%%%%%%%%%%%%%%%%%%%%%%%%%%%%%%%%%%%%%%%%%%%%%%%%%%%%%%%%%%%%%%%%%%%%%%%%%%
\appendix{\begin{center}{\bf APPENDIX B: The Special Case of the Gauge Group $\mathcal G={\mathrm SU}(2)$}\end{center}}
%%%%%%%%%%%%%%%%%%%%%%%%%%%%%%%%%%%%%%%%%%%%%%%%%%%%%%%%%%%%%%%%%%%%%%%%%%%%%%
%%%%%%%%%%%%%%%%%%%%%%%%%%%%%%%%%%%%%%%%%%%%%%%%%%%%%%%%%%%%%%%%%%%%%%%%%%%%%
%*************************************************************************
\lb{appB}
\setcounter{equation}{0}
\setcounter{lemma}{0}
\setcounter{rema}{0}
\setcounter{thm}{0}
\renewcommand{\theequation}{B\arabic{equation}}
\renewcommand{\thelemma}{B\arabic{lemma}$\:$}
\renewcommand{\therema}{{\em{\bf B\arabic{rema}$\:$}}}\vspace{.5cm}
%%%%%%%%%%%%%%%%%%%%%%%%%%%%%%%%%%%%%%%%%%%%%%%%%%%%%%%%%%%%%%%%%%%%%%%%%%%%
%%%%%%%%%%%%%%%%%%%%%%%%%%%%%%%%%%%%%%%%%%%%%%%%%%%%%%%%%%%%%%%%%%%%%%%%%%%
Here, we treat the special case of the gauge group $\mathcal G\,=\,{\mathrm SU}(2)$. We use the gluon  parametrization of the group and prove stability bounds on $Z^u_{Y,\Lambda,a}$. Before obtaining these bounds, we give some general facts about ${\mathrm SU}(2)$ and its relation to its Lie algebra ${\mathrm su}(2)$ \cite{Hall,Far}. In the gluon parametrization of the Lie algebra, we give some new results on the inverse of the exponential map, that is, the map from the Lie algebra to the group . We also show that the single plaquette, single bond variable partition function integral, expressed in terms of gluon fields, is the same as the Weyl angular eigenvalue integral.

The group ${\mathrm SU}(2)$ can be identified with the sphere $S^3$. With the $2\times 2$ identity matrix $I$ and the usual traceless and self-adjoint Pauli spin matrices $\sigma_1\,=\,\left(\barr{cc} 0&1\\1&0\earr\right)$, $\sigma_2\,=\,\left(\barr{cc} 0&-i\\i&0\earr\right)$, $\sigma_3\,=\,\left(\barr{cc} 1&0\\0&-1\earr\right)$, and setting $\vec \sigma\,=\,(\sigma_1,\sigma_2,\sigma_3)$, we write an arbitrary element as
$$
w\,=\,w_0I\,+\,i\,\vec\sigma.\vec w\,=\,\left(\barr{cc}w_0+iw_3\;&w_2+iw_1\\-w_2+iw_1\;&w_0-iw_3\earr\right)\qquad;\qquad \sum_{k=0}^3\,w_k^2\,=\,1\,.
$$
With a short hand notation, we represent the $w$ parameters as $(w_0,\vec w)\in\mathbb R^4$.

The parametrization of ${\mathrm SU}(2)$ by gluon fields in the Lie algebra ${\mathrm su}(2)$ has a desirable property. Namely, the domain of ${\mathrm su}(2)$ where the exponential map from the Lie algebra to the group is $1-1$ and the range covers ${\mathrm SU}(2)$ (with the exception of the element $-1$, i.e. minus the identity) has a simple geometric characterization. To see this, we write an arbitrary element of the Lie algebra ${\mathrm su}(2)$ as
$$
X\,=\,i\,\sum_{j=1}^3\,A_j\sigma_j\,\equiv\,i\vec A.\vec\sigma\,,
$$
where $\vec A\,=\,(A_1,A_2,A_3)\in \mathbb R^3$.

The exponential map for all $\vec A\in {\mathrm su}(2)$, noting that $(\vec A.\vec\sigma)^2\,=\,\vec A.\vec A\,=\,|\vec A|^2$, is given by
$$
e^X\,=\,e^{i\vec A.\vec\sigma}\,=\,\cos |\vec A|\,I\,+\,i\,\vec A.\vec\sigma\,\dfrac {\sin|\vec A|}{|\vec A|}\,,$$
where we adopt the power series definitions $\cos |\vec A|\,=\,\sum_{n\geq 0}\,(-1)^n\,|\vec A|^{2n}/(2n)!$ and $[\sin |\vec A|/|\vec A|\,]\,=\,\sum_{n\geq 0}\,(-1)^n\,|\vec A|^{2n}/(2n+1)!$. These are even and real analytic functions of each component $A_j$ of $\vec A$ and, of course, $[\sin|\vec A|/|\vec A|\,]\,=\,1$, for $\vec A\,=\,0$.

We represent the group element in the range by
$$
y_0I\,+\,\vec y.\vec\sigma\,=\,\left(\barr{cc}y_0+iy_3\;&y_2+iy_1\\-y_2+iy_1\;&y_0-iy_3\earr\right)\qquad;\qquad \sum_{j=0}^3\,y_j^2\,=\,1\,,
$$
and make the identification
$$
(y_0,\vec y)\,=\,\left(\cos |\vec A|\,,\,\dfrac {\sin|\vec A|}{|\vec A|}\,\vec A\right)\,.
$$

The exponential map is a jointly real analytic function of $A_1$, $A_2$ and $A_3$. On the domain $\mathcal D\,=\,\{\vec A\in\mathbb R^3\,,\,|\vec A|<\pi\}$, we show that the exponential map is $1-1$; on the boundary $|\vec A|\,=\,\pi$ the exponential map is {\em not} $1-1$; all elements are mapped to minus the identity, i.e. $(-1)$. We also show that the exponential map with domain restricted to $\mathcal D$, the range $R(\mathcal D)$ is ${\mathrm SU}(2)$ minus $(-1)$, which we denote by ${\mathrm SU}(2)/(-1)$.

We remark that as a general result, if $|\vec A|\,<\,1$, the exponential map establishes a $1-1$ relation between an open ball at the origin in ${\mathrm su}(2)$ and an open ball in the neighborhood of the identity group element of ${\mathrm SU}(2)$. This is so as the inverse element is well-defined
by its Taylor series expansion, so that $\exp:\,X\rightarrow (w_0,\vec w)$ and $\ln:\,(w_0,\vec w)\rightarrow X\,=\,i\vec A.\vec\sigma$, with $$X\,=\,i\vec A.\vec\sigma\,=\,\ln \left(w_0\,+\,i\vec w.\vec\sigma\right)
\,=\,\ln \left[1\,+\,\left(w_0\,+\,i\vec w.\vec\sigma\,-\,1\right)\right]\,.$$

If $|w_0\,+\,i\vec w.\vec\sigma\,-\,1|\,<\,1$, the matrix series $$i\vec A.\vec\sigma\,=\,\sum_{n\geq 0}\,(-1)^{n+1}\,[(w_0\,+\,i\vec w.\vec\sigma\,-\,1)^n/n]\,,$$ converges. Since
$$
|w_0\,+\,i\vec w.\vec\sigma\,-\,1|\,\leq\,|w_0\,-\,1|\,+\,|\vec w|\,=\,|\sqrt{1\,-\,\vec w^2}\,-\,1|\,+\,|\vec w|\,,
$$
convergence is guaranteed for $|\vec w|\,<\,1/2$. Also, since the $j$-th component of $\vec A$ is $A_j\,=\,-(i/2)\,{\mathrm Tr}(\sigma_jX)$, each component of $\vec A$ is a jointly real analytic function of $w_1$, $w_2$ and $w_3$.

To see that the map is $1-1$ on $\mathcal D$, for $\vec A\,,\,\vec A^\prime\in\mathcal D$, suppose that
$$
(y_0,\vec y)\,=\,\left(\cos |\vec A|\,,\,\dfrac {\sin|\vec A|}{|\vec A|}\,\vec A\right)\,=\,\left(\cos |\vec A^\prime|\,,\,\dfrac {\sin|\vec A^\prime|}{|\vec A^\prime|}\,\vec A^\prime\right)\,.
$$
Since the function $\cos u$ in strictly decreasing in $u\in[0,\pi]$, from the first component equality $\cos |\vec A|\,=\,\cos |\vec A^\prime|$ it follows that $|\vec A|\,=\,|\vec A^\prime|$. Similarly, since $\sin|\vec A|/|\vec A|\,\not=\,0$ in $\mathcal D$, we have $\vec A\,=\,\vec A^\prime$.

To prove that the range $R(\mathcal D)$ is ${\mathrm SU}(2)/(-1)$, we notice that the exponential map from $\vec A\in\mathbb R^3$ to $(y_0,\vec y)\in {\mathrm SU}(2)$ is proportional to $\vec A$. For an arbitrary element $(w_0,\vec w)$ of ${\mathrm SU}(2)/(-1)$, we have to show there exists a $\vec B\in\mathcal D\subset {\mathrm su}(2)$ such that
\bequ\lb{ball}
e^{i\vec B.\vec \sigma}\,=\,(w_0,\vec w)\,.
\eequ
For this, we consider  a general $\vec B\,=\,f(w_0,|w|)\,\vec w$ and determine the function $f$ so that Eq. (\ref{ball}) holds. Suppressing the arguments of $f$, $f\,\geq\,0$, we have
\bequ\lb{expi}
e^{i\vec B.\vec \sigma}\,=\, \cos (f|\vec w|)\,+\,i\vec w.\vec\sigma\;\dfrac{\sin (f|\vec w|)}{|\vec w|}\,=\,w_0\,+\,i\vec w.\vec\sigma\,.
\eequ
Equating the components of $\vec w$, we obtain
$$
\dfrac{\sin (f|\vec w|)}{|\vec w|}\;\vec w\,=\,\vec w\,.
$$
Next, take $f\,=\,(\sin^{-1}|\vec w|)/|\vec w|$ and set $\sin^{-1}|\vec w|\,=\,u$. Then, for fixed $|\vec w|$, the variable $u$ can take two values, namely, $u_<$, with $u_<\,<\,\pi/2$, and $u_>\,=\,\pi\,-\,u_<$. With this in mind, we have
$$
f\,=\,\dfrac1{|\vec w|}\,u\qquad;\qquad \sin(f|\vec w|)\,=\,\sin u\,=\,\sin(\pi-u)\,.
$$
Hence, equating the identity component gives the condition $\cos(f|\vec w|)\,=\,w_0$.

Now, we show that the two values of $u$ correspond to the opposite signs of $w_0$. First, for $0<u<\pi/2$, we have
$$
w_0\,=\,\cos(f|\vec w|)\,=\,\cos(\sin^{-1}|\vec w|)\,=\,(1\,-\,\vec w^2)^{1/2}\,>\,0\,.
$$
Taking now $\pi/2\,<\,u\,<\,\pi$, we have
$$
w_0\,=\,\cos(f|\vec w|)\,=\,\cos(\sin^{-1}|\vec w|)\,=\,-(1\,-\,\vec w^2)^{1/2}\,<\,0\,.
$$

Thus, both for $w_0\,>\,0$ and $w_0\,<\,0$, it follows that
\bequ\lb{vovo}
\vec B\,=\,f\,\vec w\,=\,\dfrac{\sin^{-1}|\vec w|}{|\vec w|}\;\vec w\,.
\eequ
Also, $|\vec B|\,<\,\pi$ so that $\vec B\in\mathcal D$. For $w_0=0$, $\vec w\,=\,(2/\pi)\,\vec B$.

We now determine the smoothness properties of the logarithm map.
For $w_0\,>\,0$, using the power series expansion
$$
\sin^{-1}x\,=\,x\,+\,\dfrac12\;\dfrac{x^3}3\,+\,\dfrac{1.3}{2.4}\;\dfrac{x^5}5\,+\,\dfrac{1.3.5}{2.4.6}\;\dfrac{x^7}7\,+\,\ldots\qquad;\qquad |x|\,<\,1\,,
$$
where, for $n\in\mathbb N$, $n\geq 2$, $n!!\,=\,n(n-2)!!\,=\,n(n-2)(n-4)!!\,= \,\ldots$ and $1!!\,=\,0!!\,=\,(-1)!!\,=\,1$]. We have
$\vec B\,=\,u_<\,\vec w/|\vec w|$. Hence, we have
$$
\vec B\,=\,\left[\dis\sum_{n\geq 0}\,\dfrac{(2n-1)!!}{(2n)!!}\;\dfrac{\vec w^{2n}}{2n+1}\right]\,\vec w\qquad;\qquad 0\,<\,w_0\,\leq\,1\,.
$$
Notice that the function in the square brackets is an even function of $\vec w$ and each term of series is positive. From this observation, and since $\lim_{x\rightarrow 1}\sin^{-1}x\,=\,\pi/2$, the above series for $vec B$ extends to $\vec B\,=\,(\pi/2)\,{\vec w}$, for $w_0\,=\,0$.

The above shows that the logarithm map from ${\mathrm SU}(2)$ minus $w_0\leq 0$ to $\{\vec A\in\mathbb R^3\,;\,|\vec A|<\pi/2\}$ is analytic in $w_1$, $w_2$ and $w_3$.

For $w_0\,<\,0$, $\vec B\,=\,(\pi-u_<)\,\vec w/|\vec w|$. Then,
$$
\vec B\,=\,\left[\pi\,-\,\dis\sum_{n\geq 0}\,\dfrac{(2n-1)!!}{(2n)!!}\;\dfrac{|\vec w|^{2n+1}}{2n+1}\right]\,\dfrac{\vec w}{|\vec w|}\qquad;\qquad -1\,<\,w_0\,<\,0\,.
$$
This result also extends to $w_0\,=\,0$, giving $\vec B\,=\,(\pi/2)\,\vec w$.

We now show the breakdown of analyticity of the inverse map, the logarithm map, at $w_0\,=\,0$.
For simplicity, take $\vec A\,=\,(0,0,A_3)$. Then $(w_0,\vec w)\,=\,(\cos A_3,0,0,\sin A_3)$, and $A_3\,=\,0,\pi/2,\pi$ are, respectively, mapped to the identity, $(0,0,0,1)$ and minus the identity. As $A_3$ increases from $0$ to $\pi$, the group element traces out a semi-circle in the positive $w_0-w_3$ half plane. For $A_3$ near $\pi/2$ and $w_0$ near zero, we introduce the coordinates $x$ and $y$ by letting $w_3\,=\,1\,-\,x$, $x\geq 0$, and $A_3\,=\,(\pi/2)\,+\,y$. In these coordinates, the exponential map takes $A_3$ to $w_3\,=\,\sin A_3\,=\,\cos y$ and $A_3$ to $w_0\,=\,\cos A_3\,=\,-\sin y$ or, for short,
$$
y\,\rightarrow\,(w_0=-\sin y,0,0,w_3\,=\,\cos y\,=\,1-x)\,,
$$
where negative (positive) $y$ corresponds to positive (negative) $w_0$. Here, $w_0\,=\,-{\mathrm sgn}(y)\;\sqrt{1-(1-x^2)}$, where ${\mathrm sgn}\,=\,d|x|/dx$, $x\not=0$, is the sign function.

For the logarithm map, we want $y$ as a function of $x$. Approximately,
$\cos y\,=\,1\,-\,x\,\simeq\,1-(y^2/2)$, so that $y\,=\,-2\sqrt{x}$ ($\sqrt{x}$) for $w_0>0$ ($w_0<0$), which is {\em not} real analytic at $x=0$. The exact solution, using the identity $1-\cos y\,=\,2\sin^2y/2)$, is
$$
y\,=\, -2\,{\mathrm sgn}(w_0)\,\sin\sqrt{x/2}\,=\,-2\,{\mathrm sgn}(w_0)\,\left[\sqrt{\dfrac x2}\,-\,\dfrac16\,\left(\sqrt{\dfrac x2}\right)^3\,+\,\ldots \right]\,,
$$
which shows the loss of real analyticity of the inverse, logarithm map at $w_0\,=\,0$.

\begin{rema}
	Our analysis clearly shows that the maximal domain, connected to the identity, of analyticity of the inverse, logarithm map is $0\,<\,w_0\,\leq\, 1$, or the domain of the exponential restricted to $|\vec A|\,<\,\pi/2$. Unfortunately, the Brouwer invariance domain \cite{Brou} theorem only applies for $1-1$ and continuous maps in from open domains in $\mathbb R^n$ to an open  in $\mathbb R^n$ (not from $\mathbb R^n$ to $\mathbb R^m$, $m\not= n$, as here!) Whenever $m=n$ the theorem states that the inverse map is also continuous on the map range. However, for $B_{<\alpha}\,\equiv\,\{\vec x\in\mathbb R^3; 0\leq |\vec x|<\alpha \}$, the map $h$ from $B_{<\pi/2}$ to $B_{<1}$ is given by $\vec w\,=\,h(\vec B)\,=\,\sin |\vec B|\,\vec B/|\vec B|$ is continuous and $1-1$ so that, by the invariance domain theorem, the inverse is continuous. In this case, the inverse $h^{-1}$ is explicitly given by $h^{-1}(\vec w)\,=\,\sin^{-1}|\vec w|\,\vec w/|\vec w|$. We easily verify that $(h\circ h^{-1})(\vec w)\,=\,\vec w$.
\end{rema}

The above approach to show surjectivity of the exponential map and to determine the logarithm is geometric in flavor. We can also use a spectral approach to determine the logarithm.
	
In this approach, which also applies to $\mathrm U(N)$ and ${\mathrm SU}(N)$, we use spectral theory for the unitary matrix $U$. In the special case $w=(w_0,\vec w)\in {\mathrm SU}(2)$, the same formula for the logarithm results. We briefly describe the method for this case. Consider the following commutative diagram where $V$ is a unitary matrix that diagonalizes $U\not= 0$, i.e. $$A\,=\,V^{-1}UV\,=\,{\mathrm diag}(e^{i\la},e^{-i\la})\,\equiv\,e^{iB}\qquad;\qquad \la\in(-\pi,\pi]\,,$$
since ${\mathrm Tr}\,B\,=\,0$,
$$\barr{ccc}
(w_0,\vec w)\,=\,U&\xlongrightarrow{\:\:\:\:\:\:\:\:\:\:\:\:V\:\:\:\:\:\:\:\:\:\:\:\:}&V^{-1}UV\,=\,e^{iB}\vspace{2mm}\\
\exp\bigg\uparrow&&\bigg\uparrow{\exp}\vspace{2mm}\\
iX&\xlongrightarrow{\:\:\:\:\:\:\:\:\:\:\:\:V\:\:\:\:\:\:\:\:\:\:\:\:}&iV^{-1}XV\,=\,{iB}\:\:.
\earr$$
	
To show that the exponential map is surjective in some domain, we need $X$. However, in the diagonal form of the right-hand-side above, $B\,=\,{\mathrm diag}(\la,-\la)$ so that $X\,=\,VBV^{-1}$. Now, $U$ has an orthonormal basis of eigenvectors and $V$ is the matrix of column eigenvectors. For ${\mathrm SU}(2)$, the eigenvalues of $U$ are $e^{\pm i\la}\,=\,w_0\,\pm\,i|\vec w|$ and $\pm\la\,=\,-i\ln (w_0\,\pm\,i|\vec w|)$. The normalized eigenvectors are, for $w_0\,\geq 0$, $v_1\,=\,(1/\mathcal M,b/\mathcal M)$ and $v_2\,=\,(-\bar b/\mathcal M,1/\mathcal M)$, where
$$b\,=\,\dfrac{-i(w_2-|\vec w|)}{w_2+iw_1}\qquad;\qquad \mathcal M\,=\,\left[\dfrac{2|\vec w|(|\vec w|-w_3)}{w_1^2\,+\,w_2^2}\right]^{1/2} \,.$$ The column eigenvectors $v_1$ and $v_2$ are the columns of the matrix $V$.

Calculating $X\,=\,VBV^{-1}$ gives the same result as Eq.(\ref{vovo}).
	
It is helpful to note that
$$
i\la\,=\,\ln[w_0\,+\,i|\vec w|]\,=\,\ln\left[\left(w_0^2\,+\,\vec w^2\right)^{1/2}\,e^{i\alpha} \right]\,=\,i\alpha\,,
$$
with $\tan \alpha\,=\,|\vec w|/w_0$, $0\,\leq\,\alpha\,<\,\pi/2$, and $\cos\alpha\,=\,w_0$.

\begin{rema}
We emphasize that, for the groups ${\mathrm SU}(N\geq 3)$ to find a parametrization and a minimal domain $\mathcal D$ of the parameters on which the exponential map is $1-1$ and onto \cite{Simon2,Hall} (or, at least, dense in the group) is a much more difficult task then for the case ${\mathrm SU}(2)$ we treat here.
\end{rema}
	
Next, we give the ${\mathrm SU}(2)$ Haar measure. With $\mathcal N=2\pi^2$, we have \cite{GJ,Simon2}]
\bequ\lb{haarsu2}
d\nu(\vec A)\,=\,\rho(|\vec A|)\,dA_1dA_2dA_3\quad;\quad \rho(|\vec A|)\,=\,\dfrac 1{\mathcal N}\,\dfrac {\sin^2|\vec A|}{|\vec A|^2}\,,
\eequ
which can be  easily bounded from above and below. The Haar measure of the set $|\vec A|=\pi$ of the domain $\mathcal D$ is zero. The same holds, of course, for the single group element $(-1)$.

The Haar measure $d\nu(\vec A)$ of Eq. (\ref{haarsu2}) is obtained in a geometrical way by considering the correspondence between $w\,=\,w_0 I\,+\,i\vec\sigma\cdot\vec w$ in ${\mathrm SU}(2)$ and $S^3$ with coordinates $(w_0,\vec w)$, with the constraint $\sum_{k=0,\ldots,3}\,w_k^2\,=\,1$. Multiplication of a group element $w$ by an arbitrary group element $u$ corresponds to an orthogonal transformation in $\mathbb R^4$. The transformation preserves the measure. Using spherical coordinates in $\mathbb R^4$, calculating the Jacobian and restricting to $S^3$ gives the formula for the Haar measure, up to a normalization.

We now give the key point that is used to establish the relation between the gluon parametrized partition function integral and the Weyl angular eigenvalue integral. In the gluon parametrization of ${\mathrm SU}(2)$, we write an element as $U\,=\,e^{i\vec\sigma\cdot\vec A}$. Since $\vec\sigma\dot\vec A$ is self-adjoint and $(\vec\sigma\cdot\vec A)^2\,=\,\vec A\cdot\vec A$, the $2\times 2$ matrix $\vec\sigma\cdot\vec A$ has eigenvalues $\pm|\vec A|$, which are precisely the angular eigenvalues of $U$.
	
Now, we obtain the stability bounds for $\mathcal G\,=\,{\mathrm SU}(2)$ in Eq. (\ref{Zuw}), we fix the enhanced temporal gauge. Using the above considerations, Eq. (\ref{expi}) and proceeding as in the proof of the upper stability bound as in the general case of a compact and connected Lie gauge group $\mathcal G$, we have
$$Z^w_{Y,\Lambda,a}\,\leq\,z^{\Lambda_r}\,,$$
where $d\nu(\vec A)$ given in Eq. (\ref{haarsu2}),
$$
z\,=\,\dis\int_{|\vec A|<\pi}\,\exp\left\{-\frac{a^{d-4}}{g^2}\,S(A)\right\}\,d\nu(\vec A)\,.
$$
where $S(A)\,=\,\|1-e^{i\vec\sigma\cdot\vec A}\|^2_{H-S}
\,=\,2\,{\mathrm Tr}(I-\cos|\vec A| I)\,=\,4\,(1-\cos|\vec A)$.

Passing to spherical coordinates in $\mathbb R^3$, i.e. $r=|\vec A|$, and $\Omega$ being the solid angle, $|\Omega|\,=\,4\pi$, we get
$$
z\,=\,\dfrac1{2\pi^2}\,|\Omega|\,\dis\int_0^\pi e^{-\left( 4a^{d-4}/g^2\right)\,(1-\cos r)}\,\sin^2r\,dr
$$

we now turn to the Weyl integration formula. For the gauge group $\mathrm U(N)$ and a class function $f(\lambda)\,\equiv\,f(\lambda_1,\ldots,N)$ of the angular eigenvalues of the unitary $U$, $\lambda_1,\ldots,\lambda_N\in(-\pi,\pi]$, $U$ is unitarily equivalent to the matrix ${\mathrm diag}(e^{i\lambda_1},\ldots,e^{i\lambda_N})$, and the Weyl formula is
$$
\dis\int_{{\mathrm U}(N)}\,f(U)\,d\mu(U)\,=\,\dfrac 1{N!(2\pi)^N}\,\dis\int_{(-\pi,\pi]^N)}\, f(\lambda)\,\rho(\lambda)\,d^N\lambda\,,
$$
where $\rho(\lambda)=\prod_{1\leq j<k\leq N}\,\left|
e^{i\lambda_j}\,-\,e^{i\lambda_k}\right|^2$.

For the Wilson action and ${\mathrm U}(2)$,
$$
f(\lambda)\,=\,\exp\left[-\dfrac{a^{d-4}}{g^2}\,\|1-U \|^2_{H-S}\right] \,=\,
\exp\left[-\dfrac{a^{d-4}}{g^2}\,2\sum_{j=1,2}(1-\cos \lambda_j)\right]\,,
$$
where we used that $U$ is unitarily equivalent to ${\mathrm diag}\left(e^{i\lambda_1},e^{i\lambda_2} \right)$.

To obtain the Weyl formula for ${\mathrm SU}(2)$ from the one for $\mathrm U(2)$ (see \cite{Simon2}), we just have to insert a $[2\pi\delta(\lambda_1+\lambda_2)]$ constraint factor in the above integrand and e.g. carry out the $\lambda_2$ integration. Doing this, for ${\mathrm SU}(2)$, we obtain
$$
z\,=\,\dfrac 1{4\pi}\,\dis\int_{(-\pi,\pi])}\,\rho(\lambda_1)\,e^{-\dfrac{a^{d-4}}{g^2}\,S(\lambda_1)}\,d\lambda_1\,,
$$
where $S(\lambda_1)\,=\,4(1-\cos\lambda_1)$ and $\rho(\lambda_1)\,=\,4\sin^2\lambda_1$. Finally, since the above integrand is even, we get
$$
z\,=\,\dfrac 2{\pi}\,\dis\int_{(0,\pi]}\,\rho(\lambda_1)\,e^{-\frac{a^{d-4}}{g^2}(1-\cos\lambda_1)}\,\sin^2\lambda_1\,d\lambda_1\,,
$$
which is the same as the gluon integral.

We now prove the upper bound on $z$. Using, for all $\vec A$ in the integration domain, $$\|1-e^{i\vec\sigma.\vec A}\|^2_{H-S}\,=\,2\left(1\,-\,\cos|\vec A|\right)\,\geq\,\dfrac4{\pi^2}\,|\vec A|^2\,,$$
and also $\rho(\vec A)\,\leq\,1/\mathcal N$, for all $\vec A$, we get, using $(\sin u/u)\leq 1$, $u>0$,
$$
z\,\leq\,\dfrac1{\mathcal N}\;\dis\int_{|\vec A|<\pi}\,\exp\left[-\frac{4a^{d-4}}{\pi^2g^2}\,|\vec A|^2\right]\,\dfrac{\sin^2|\vec \vec A|}{|\vec A|^2}\,d\vec A\,\leq\,\dfrac{4\pi}{\mathcal N}\,\dis\int_0^\pi\,\exp\left[-\frac{4a^{d-4}}{\pi^2g^2}A^2\right]\,A^2\,dA\,.
$$
Making the change of variables $r\,=\,\dfrac{2a^{(d-4)/2}}{\pi g}A$ , gives, with $\gamma\,=\,\dfrac{2a^{(d-4)/2}}{g}$,
$$
z\,\leq\,\left(\dfrac {g^2}{a^{d-4}}\right)^{3/2} \dfrac{\pi^2}4\,E(\gamma)\,\leq\,\left(\dfrac {g^2}{a^{d-4}}\right)^{3/2}\, \dfrac{\pi^2}4\,E(\infty)\,,
$$
where \bequ\lb{EE}E(\gamma)\,=\,\int_{r\leq\gamma}\,e^{-r^2}\,r^2dr\,.\eequ
	
In this way, noting that $E(\gamma)\,\leq\,E(\infty)=\pi/2$,we have the upper stability bound
$$
Z^w_{Y,\Lambda,a}\,\leq\,\left(\dfrac {g^2}{a^{d-4}}\right)^{3\Lambda_r/2}\, \left[\dfrac{\pi^3}8\right]^{\Lambda_r}\,.
$$
	
We proceed similarly for the lower bound. We first observe that $\mathcal A_p$ in the (four-bond) plaquette action $(a^{d-4}/g^2)\,\mathcal A_p$ satisfies the global quadratic upper bound, with $C\,=\,2N^{1/2}$,
$$
\mathcal A_p\,\leq\,C^2\,\sum_{1\leq j\leq 4}|\vec A_{p,j}|^2\,.
$$
After reducing the integration domain by half, from $|\vec A|\,<\,\pi$ to $|\vec A|\,\leq\,\pi/2$, a lower bound on the density $\rho(\vec A)$ results from using $[\sin(|\vec A|)/|\vec A|]\,\geq (2/\pi)$ on the reduced domain. From this, we have
$$
Z^w_{Y,\Lambda,a}\,\geq\,\dis\int_{|\vec A|\leq\pi/2}\, \exp\left[-\frac{a^{d-4}C^2}{g^2}\,\sum_p\sum_{1\leq j\leq 4}|\vec A_{p,j}|^2\right]\,\prod_{b}\left[\dfrac4{\mathcal N\pi^2}\,\,d\vec A_b\right]\,.
$$
	
At this stage, the integral factorizes over bonds. Since each lattice bond belongs to a maximum of $2(d-1)$ distinct plaquettes, paying with this factor, we change the sum over $p$ to a sum over bonds. In doing this, we lower the lower bound, the integral factors over the bonds and we get
$$
Z^w_{Y,\Lambda,a}\,\geq\,\tilde z^{\Lambda_r}\,,
$$
where
$$
\tilde z\,=\,\dfrac2{\mathcal N\pi^2}\,\dis\int_{|\vec A_b|\leq\pi/2}\, \exp\left[-\frac{2a^{d-4}(d-1)C^2}{g^2}\,|\vec A_b|^2\right]\,d\vec A_b\,.
$$
	
Passing to spherical coordinates $(A=|\vec A|,\theta,\phi)$ and then making the change of variables $r\,=\,\left[\dfrac{a^{(d-4)/2}C[2(d-1)]^{1/2}}{g}\right]A\,$
gives
$$
\tilde z\,=\,\left[\dfrac{g^2}{a^{d-4}}\right]^{3/2}\,\left(\dfrac2{\pi C[2(d-1)]^{1/2}}\right)^3\,E(a^{(d-4)/2}\pi C[2(d-1)]^{1/2}/2g)\,.
$$
	
We remark that $E(u)$ given in Eq. (\ref{EE}) is monotone increasing. Its maximum occurs at $a=1$ and the largest value $g_0$ we take for $g$. We denote the integral with these values of parameters by $E_0$. Hence we get the lower stability bound
$$
Z^w_{Y,\Lambda,a}\,\geq\,\left[\dfrac{g^2}{a^{d-4}}\right]^{3\Lambda_r/2}\,\,\left\{ \left(\dfrac2{\pi C[2(d-1)]^{1/2}}\right)^3\,E_0\right\}^{\Lambda_r}\,.
$$
	
Thus, extracting the same singular factor $\left[\dfrac{g^2}{a^{d-4}}\right]^{3\Lambda_r/2}$ from the upper and lower stability bounds, the scaled partition function defined by $Z_{Y,\Lambda,a}\,=\,\left[g^2/a^{d-4}\right]^{-3\Lambda_r/2}\,Z^w_{Y,\Lambda,a}$ obeys thermodynamic and ultraviolet stability bounds, as given in Theorem \ref{thm2}.
\begin{acknowledgements}We thank FAPESP and CNPq for support, and Profs. Tiago Pereira and Marcel Novaes for discussions on random matrix theory.\end{acknowledgements}

\end{document}